\documentclass[aps,prd,showpacs,nobibnotes,twocolumn,preprintnumbers,
nobalancelastpage,superscriptaddress]{revtex4-1}
\usepackage{amssymb}
\usepackage{amsmath}
\usepackage{graphicx}
\usepackage{latexsym}
\usepackage{dcolumn}
\usepackage{bm}
\input{epsf}
\newcommand{\be}{\begin{equation}}
\newcommand{\ee}{\end{equation}}
\newcommand{\ba}{\begin{eqnarray}}
\newcommand{\ea}{\end{eqnarray}}
\newcommand{\no}{\nonumber}
\newcommand{\bfi}{\begin{figure}
\epsfxsize=8cm
\epsffile}
\newcommand{\bfig}{\begin{figure*}
\epsfxsize=18cm
\epsffile}
\newcommand{\efi}{\end{figure}}
\newcommand{\efig}{\end{figure*}}
\newcommand{\bi}{\begin{itemize}}
\newcommand{\ei}{\end{itemize}}
\newcommand{\la}{\lesssim}
\newcommand{\ga}{\gtrsim}
\newcommand{\mpch}{h^{-1} {\rm Mpc}}
\newcommand{\gpch}{h^{-1} {\rm Gpc}}

\newcommand{\bibmnras}{Mon. Not. R. Astron. Soc. }
\newcommand{\bibapj}{Astrophys. J. }
\newcommand{\bibapjl}{Astrophys. J. Lett. }
\newcommand{\bibapjs}{Astrophys. J. S. }
\newcommand{\bibjcap}{J. Cosmol. Astropart. Phys. }
\newcommand{\bibprd}{Phys. Rev. D }

\newcommand{\bibaap}{Astron. Astrophys. }

\newcommand{\model}{\mathrm{model}}
\newcommand{\simu}{\mathrm{simu}}
\newcommand{\mock}{\mathrm{mock}}

\begin{document}

\title{Fast generation of weak lensing maps by the inverse-Gaussianization method}
\author{Yu Yu}
\email{yuyu22@shao.ac.cn}
\affiliation{Key laboratory for research in galaxies and cosmology, Shanghai Astronomical Observatory, Chinese Academy of Sciences, 80 Nandan Road, Shanghai, China, 200030}

\author{Pengjie Zhang}
\email{zhangpj@sjtu.edu.cn}
\affiliation{Center for Astronomy and Astrophysics, Department of Physics and Astronomy,
Shanghai Jiao Tong University, Shanghai, 200240}
\affiliation{IFSA Collaborative Innovation Center, Shanghai Jiao Tong University, Shanghai, China, 200240}
 
\author{Yipeng Jing}
\affiliation{Center for Astronomy and Astrophysics, Department of Physics and Astronomy,
Shanghai Jiao Tong University, Shanghai, 200240}
\affiliation{IFSA Collaborative Innovation Center, Shanghai Jiao Tong University, Shanghai, China, 200240}

\begin{abstract}
To take full advantage of the unprecedented power of 
upcoming weak lensing surveys, understanding the noise, such as cosmic
variance and geometry/mask effects, is as important as
understanding the signal itself. Accurately quantifying the
noise requires a large number of statistically independent mocks for a variety of
cosmologies. This is impractical for weak lensing simulations, which
are costly for simultaneous requirements of large box size (to cover a significant fraction of the past
light cone) and high resolution (to robustly probe the small scale
where most lensing signal resides). 
Therefore fast mock generation methods are desired and are under
intensive investigation. 
We propose a new fast weak lensing map generation method, named the inverse-Gaussianization method,
based on the finding that a lensing convergence field can be Gaussianized
to excellent accuracy by a local transformation \cite{yuyu11}. Given a
simulation, it enables us to produce up to an infinite number of statistical lensing maps as quickly as we can produce the simulatin initial conditions.
The proposed method is tested against simulations for each tomography
bin centered at lens redshift $z\sim 0.5$,
$1$, and $2$, with various statistics. We find that the lensing maps
generated by our method have reasonably accurate  power spectra,
bispectra, and power spectrum covariance matrix. Therefore, it will be useful  for weak lensing surveys
to generate realistic mocks. 
As an example of application, we measure the probability distribution function of the lensing power
spectrum,  from 16384  lensing maps produced by the
inverse-Gaussianization method. 
\end{abstract}
\pacs{}
\maketitle


\section{Introduction}
\label{sec:Introduction}
The large-scale structure (LSS) of the Universe provides us invaluable tools, such as the baryon acoustic oscillation, weak gravitational lensing, redshift space distortions, and cluster abundance,  to probe the dark universe. Various ongoing and upcoming projects, such as DES, eBOSS, HSC,  DESI, LSST, Euclid, and WFIRST, aim to measure these large-scale structures to unprecedented precision.  Over the next decade,  orders of magnitude improvement in constraints of dark energy and the nature of gravity are expected.  

However, to match the power of these massive dark energy surveys and
to constrain cosmology tightly and robustly, the requirement of
theoretical modelling of related LSS statistics is highly challenging.
In particular, not only do we need to understand the signal precisely, but we
must also understand the associated error precisely. One example is
the weak lensing power spectrum (signal) and the associated covariance
matrix (statistical error of the signal). The error estimation  is
more challenging, for several reasons. First, estimating error
involves higher-order statistics and therefore stronger nonlinearity
and non-Gaussianity
(e.g., Refs. \cite{Scoccimarro99,Rimes05,Hamilton06,Rimes06,Neyrinck06,Neyrinck07,Lee08,
  Semboloni07,Takada09,Pielorz10,Kiessling11b}).  Second, it depends
on survey boundaries and masks in complicated ways
(e.g., Ref. \cite{Harnois-Deraps15}). Both make it hard to model the error
accurately by analytical methods.  In general, mocks generated by
numerical cosmological simulations are necessary for the robust
evaluation of errors. However, given that sample variance in error is
in general larger than that in signal,  many more mocks are required
for the accurate determination of error
\cite{Dodelson13,Taylor13,Percival14}. Furthermore,  to robustly
handle the nonlinearity and non-Gaussianity, high resolutions are
required for numerical simulations
(e.g., Refs. \cite{Harnois-Deraps12,Blot15,Blot15arxiv}). Even more challenging, to robustly constrain cosmology,
mocks of various cosmologies are required. 
These requirements further increase the difficulty of mock generation.

Alternative ways have been proposed for fast generation of mocks. For 3D galaxy mocks, there are the lognormal model \cite{Coles91,Cole05}, PTHALOS \cite{Scoccimarro02,Manera13,Manera15}, PINOCCHIO \cite{Monaco02,Monaco13}, 
Quick Particle Mesh Simulations (QPM) \cite{White14}, augmented Lagrangian perturbation theory (ALPT), \cite{Kitaura13b}, EZmocks \cite{Chuang15}, COLA \cite{Tassev13,Tassev15,Koda15}, L-PICOLA\cite{Howlett15}, ICE-COLA \cite{Izard15arxiv}, HALOGEN \cite{Avila15}, FastPM \cite{Feng16}, etc. 

Weak lensing mocks have different requirements. First, they require
more accurate modelling of the nonlinear regime where the majority of
weak lensing signal is located. As a consequence,  the Lagrangian-based
methods are inappropriate due to their long-standing problem of
underestimating nonlinear powers. Second, they require larger volume
since weak lensing is a projected LSS over a $\sim c/H_0=3\gpch$
distance.   A popular choice in the literature is the lognormal model. It has been demonstrated as a useful approach to capture cosmological information \cite{Seo11,Seo12} and a valuable generator of weak lensing mocks \cite{Carron14b,Takahashi14,Carron15,Friedrich16}).  However, the lognormal model has limitations. It fails at modelling LSS
 in the low density regime (e.g., Ref. \cite{Colombi94}).  The actual probability distribution function (PDF)
 of lensing convergence shows significant deviation from lognormal \cite{yuyu11}. Other limitations are reported too \cite{Xavier16}.

In this paper, we focus on fast generation of weak lensing
maps. Observational effects can be further added to these lensing
  maps to generate mocks for cosmological surveys. This step depends
  on specifications of individual surveys and requires specific
  expertise. Therefore, throughout the paper, we will only discuss the
  generation of lensing maps, corresponding to mocks of perfect
  observations.  In the paper, we may use 
  ``mocks'' and ``lensing maps'' interchangeably. 
 We propose the ``inverse-Gaussianization'' method,  directly extending from our previous ``Gaussianization'' works \cite{yuyu11,yuyu12}. {\it Gaussianization} refers to applying a local transform to make the one-point PDF of the field Gaussian.
\citet{Scherrer10} proposed the Gaussian Copula Hypothesis to the matter field,
 which implies that the matter field will be fully Gaussianized, when only a local transform is applied to Gaussianize the one-point PDF.
Motivated by Ref. \cite{Scherrer10}, \citet{yuyu11,yuyu12} applied Gaussianization to a lensing convergence field constructed from N-body simulation and found that the Gaussianization suppresses the non-Gaussian statistics (including up to sixth-order cumulants of the smoothed fields and reduced bispectrum) a lot for the shape-noise-free convergence field.
The Gaussianized lensing convergence field is close to Gaussian, which motivates us to propose the following inverse-Gaussianization method.

The local transform function for Gaussianization is fully determined by the non-Gaussian field, 
once the irrelevant variance of the resulting field is specified.
Although the local transform function depends on cosmology, from several simulation realizations, we could  deterministically measure the ensemble average.
Applying the averaged local transform to the simulation realizations, one could measure the power spectrum of the Gaussianized fields.
According to the measured power spectrum, one could produce the Gaussian random fields (GRFs) just like one produces the initial conditions for cosmological simulations.
To generate mocks, one could inversely transform the GRFs by the measured local transform function.
We call the above process the {\it inverse-Gaussianization} method.
The resulting field would have the same statistics if the Gaussian Copula Hypothesis is perfect.
If the assumption is imperfect, at least the resulting field should be similar to the real one.

This paper presents the idea of fast mock generation by the inverse-Gaussianization method.
We describe the details of the method in Sec. \ref{sec:method}. Its performance is quantified in Sec. \ref{sec:tests}.  We find that the generated lensing maps have a reasonably accurate covariance matrix.  An application of the inverse Gaussianization is presented in Sec. \ref{sec:application}. 
We conclude and discuss in Sec. \ref{sec:conclusion} and \ref{sec:discussion}.


\section{Inverse-Gaussianization method}
\label{sec:method}

Weak lensing directly probes the matter overdensity projected along
the line of sight from the observer to the source galaxies. The
lensing convergence for a source at redshift $z_s$ and angular
position $\hat{n}$ is 
\ba
\label{eqn:kappa}
\kappa(\hat{n},z_s)=\int_0^{\chi_s}W(z,z_s)\delta(\hat{n},z)d\chi(z)\ .
\ea
Here, the lensing kernel $W(z,z_s)$ has a broad width and varies
slowly with $z$. $\chi(z)$ is the radial coordinate to redshift $z$, and
$\chi_s=\chi(z_s)$. For a flat cosmology, $\chi$ is also the comoving
angular diameter distance. 
Through the source redshift $z_s$ dependence in lensing convergence
$\kappa$, the lensing tomography works to reconstruct the 3D matter
distribution.  In reality, due to limited source redshifts,
measurement errors, and the relatively wide and smooth lensing kernel,
the reconstruction is most feasible for the binned matter overdensity
$\delta^\Sigma$.  To reasonably excellent approximation,
  Eq. (\ref{eqn:kappa}) can be written as
\ba
\label{eqn:Wi}
\kappa\simeq \sum_i W_i \delta_i^\Sigma\ .
\ea
Here, $\delta^\Sigma_i$ is the surface overdensity of the $i$th
redshift bin ($z_i-\Delta z_i<z<z_i+\Delta z_i$).  $W_i\simeq
W(\chi_i,\chi_s)\Delta \chi_i$ and $\Delta \chi_i=\chi(z_i+\Delta
z_i/2)-\chi(z_i-\Delta z_i/2)$. For a typical source redshift
$z_s=1$, the above approximation is excellent
as long as $\Delta \chi_i\la 300 h^{-1}$Mpc ($\Delta z_i\la 0.1$) and
is reasonably good as long as $\Delta z_i\la 0.2$. Thus, in this work, we focus on the
projected matter field $\delta^\Sigma$ instead of the true lensing
convergence field.

Working on fast generations of $\delta^\Sigma$ instead of
   $\kappa$ has two advantages. First,  one can  conveniently produce lensing
convergence maps to any source  redshifts , by linearly combining
$\delta^\Sigma$ generated at each (lens) redshifts. Therefore, it is more
flexible for lensing mock generations, and it is more convenient for
lensing tomography analysis.  Second, the performance test on the
$\delta^\Sigma$ field is a stronger version than the test on the $\kappa$
field.   Both the lensing convergence $\kappa$ and the projected matter field $\delta^\Sigma$ are, in principle, observable.  $\delta^\Sigma$ is a redshift resolved version of $\kappa$.  It contains more detailed information on the LSS evolution and hence the dark universe and presents stronger non-Gaussianity than $\kappa$.
The tests presented in this work mainly concern how well non-Gaussian features, like the mode coupling, could be recovered by the proposed method.
Therefore, working on $\delta^\Sigma$ instead of $\kappa$ allows us to
more reliably evaluate the proposed inverse-Gaussinization  method. 

The inverse-Gaussinization is straightforward to implement in
simulations, with the following steps:
\bi
\item[(i)] {\it Step 1: determining the Gaussianization function}.---For a given simulation snapshot
  at a given redshift, we can obtain a number of $\delta^\Sigma$
  fields by projecting along different directions with required
  projection depth and by splitting the
  simulation volume into subvolumes.  The
  Gaussianization function $y(\delta^\Sigma)$ is a local transform determined by 
\be
y(\delta^\Sigma)=\mathrm{erf}^{-1}(2\mathrm{cdf}(\delta^\Sigma)-1)\ .
\label{eqn:formfunc}
\ee
Here, $\mathrm{cdf}$ is the cumulative distribution function of the
$\delta^\Sigma$ field. $\delta^\Sigma$ is non-Gaussian. But by
definition the resulting $y$ field has a Gaussian one-point PDF, with
mean of zero and $\sigma_y=\sqrt{2}/2$. A surprising and crucial
feature of LSS is that the above local transform actually renders the
whole field (not just the one-point PDF) Gaussian, as shown in
\citet{yuyu11,yuyu12}. One can obtain individual $y(\delta^\Sigma)$
from each $\delta^\Sigma$ field and then do the average to obtain the
averaged one. Or one can first obtain the averaged cdf$(\delta^\Sigma)$
combining all $\delta^\Sigma$ fields and then obtain the averaged
$y(\delta^\Sigma)$.  In principle, the two averagings do not commute,
and the resulting $y(\delta^\Sigma)$ differ. In reality, we find no
significant difference. 
\item[(ii)] {\it Step 2: obtaining the power spectrum $C_y(\ell)$ of $y$ maps}.---Applying the
  Gaussianization to each $\delta^\Sigma$ map, we obtain the
  corresponding $y$ maps. We can then measure their power spectra to
  determine the averaged one. 
\item[(iii)] {\it Step 3: Generating realizations of the $y$ fields}.--- Since the
  starting point of the inverse Gaussianization is that $y$ is
  Gaussian \cite{yuyu11,yuyu12}, its statistics is completely
  determined by the power spectrum $C_y(\ell)$. Therefore, given
  $C_y(\ell)$ measured from step 2, we can generate many (infinite, to
  be exact) independent realizations. This is done by convolving with
  a Gaussian random field with a flat power spectrum (white noise). 
\item[(iv)]  {\it Step 4: generating realizations of $\delta^\Sigma$ fields}.---Using
  the Gaussianization function ($y$-$\delta^\Sigma$ relation) obtained
  in step 2,  we inversely transform the $y$ maps 
  generated in step 3 to obtain realizations of the $\delta^\Sigma$
  fields. All of these realizations are independent of each other, and
  we can generate as many of such realizations as we need. 
\item[(v)] {\it Step 5: generating weak lensing maps}.---Linearly combining
  $\delta^\Sigma$ maps generated at each redshift with appropriate
  weighting [Eq. (\ref{eqn:Wi})], we will obtain weak lensing maps to a given
  redshift. This last step is trivial for the purpose of this
  paper. Therefore, we will stop at step 4. 
\ei
This inverse-Gaussianization method is similar to producing mocks adopting lognormal approximation,
 in the sense of adopting a local transform to the GRFs.
One may find that all the steps above are deterministic. 
All the ingredients we need to produce mocks are measured from several simulation realizations.
No parameter is introduced in the whole pipeline.

This inverse-Gaussianization method directly produces a 2D projected density field over thickness of $300\mpch$, similar to lognormal models.
Due to this fact, it cannot recover the detailed 3D information.
Compared to the method adopting a fast simulation technique like QPM, it cannot produce a consistent halo distribution.
Thus, the application of these mocks produced by the inverse-Gaussianization method is limited to the studies not involving halos. 


\section{Performance tests}
\label{sec:tests}
The inverse-Gaussianization procedure requires verification against simulations, since the Gaussianization
of the weak lensing field cannot be done perfectly \cite{yuyu11}. Namely,
the $y$ field after the local Gaussianization [Eq. (\ref{eqn:formfunc})]
contains weak, nevertheless detectable, non-Gaussianity globally
\cite{yuyu11}. Therefore, the inverse-Gaussianization method is not
strict, and its performance shall be quantified against
simulations. This section tests it through a number of
tests.

\subsection{Simulation and $\delta^\Sigma$ fields}
\label{sec:simulation}

The simulation we use to test the inverse-Gaussianization method is
run by a particle-particle-particle-mesh 
code,  detailed in \citet{jingyp07} . The simulation has  box size of $1.2\gpch$ and particle number of
$3072^3$. 
A flat $\Lambda$CDM cosmology is adopted with $\Omega_m=0.268$,
$\Omega_\Lambda=0.732$, $\sigma_8=0.83$ and $n_s=0.968$. 
We cut the simulation box into curved slices to make two-dimensional weak lensing convergence maps.
First, we divide the box into four cuboids with a side length of $1200\mpch$
along the line of sight and $600\mpch$ along the transverse direction.
For each cuboid, along the line of sight we could make four curved slices of $300\mpch$ thickness adopting the periodical boundary condition.
Considering projection along three dimensions, we totally make 48 lensing convergence realizations from one simulation box.
It has been proven by \citet{panjun05} and \citet{Hartlap07} that if the number of realizations is less than the number of band power bins, the sample covariance is singular.
As a remedy, \citet{panjun05} proposed to use singular value decomposition (SVD) to obtain the pseudoinverse.
However, \citet{Hartlap07} strongly discouraged the use of the SVD since the large bias is induced in the estimation of the precision matrix.
Nevertheless, 48 realizations enable us to do analysis for the band power measurement on 20 $\ell$ bins.

The cosmic microwave background (CMB)-lensing kernel peaks at $z\sim 2$.
Current lensing surveys, with typical source redshift $z_s\sim 1$,
probe the matter distribution at $z\gtrsim 0.5$. 
Future lensing surveys can be much deeper, enabling us to probe the matter distribution at $z\sim 1$.
Thus, we choose the snapshots of redshift $z=2.023$, $1.028$, and
$0.485$ to cover typical lens redshifts that can be probed by weak
gravitational  lensing. 
For two low redshift slices at $z=1.028$ and $0.485$, we cut the map to size of $13\times 13 \deg^2$, 
while for the high redshift slices, we set the map size to $8.8 \times 8.8 \deg^2$.
Since the lensing kernel is a broad and slowly varying function around the intermediate redshift,
we directly project the slices with $300\mpch$ thickness into the two-dimensional $\delta^\Sigma$ field.
The grid is set to $512\times512$ uniform grid, which corresponds to a pixel size of $1.03'$ for $z=2.023$ and $1.52'$ for $z=1.028$ and $0.485$.

\begin{figure}
\epsfxsize=8cm
\epsffile{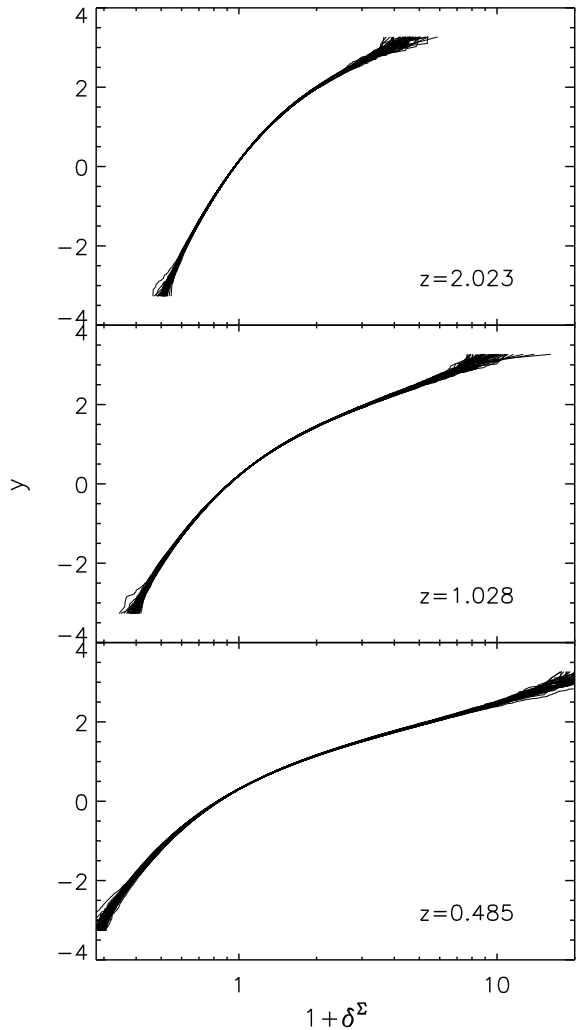}
\caption{The local transforms $y(\delta^\Sigma)$ obtained from 48
  realizations for each redshift, and after the transformation, the y
  field has a Gaussian PDF. The local transforms diverse at both the
  high- and low-density ends, reflecting the rareness of the very high
  density and very low density regions. They correspond to {\bf Step
    1} of the inverse-Gaussianization method, detailed in Sec. \ref{sec:method}. }
\label{fig:transforms}
\end{figure}

\subsection{Local transforms}
\label{sec:transforms}

The local transforms obtained from 48 simulation realizations are
presented in Fig. \ref{fig:transforms}. 
The results for $z=2.023$, $1.028$, and $0.485$ are presented from top to bottom, respectively.
We present the projected density field in the form of
$1+\delta^\Sigma$ in logarithmic scale. 
We find good convergence for the local transform functions in the intermediate density regime.
Divergence only appears at both the high and low ends.
At these regimes, the local transform function is dominated by the extreme values of the field, which experience large cosmic variance.

\bfig{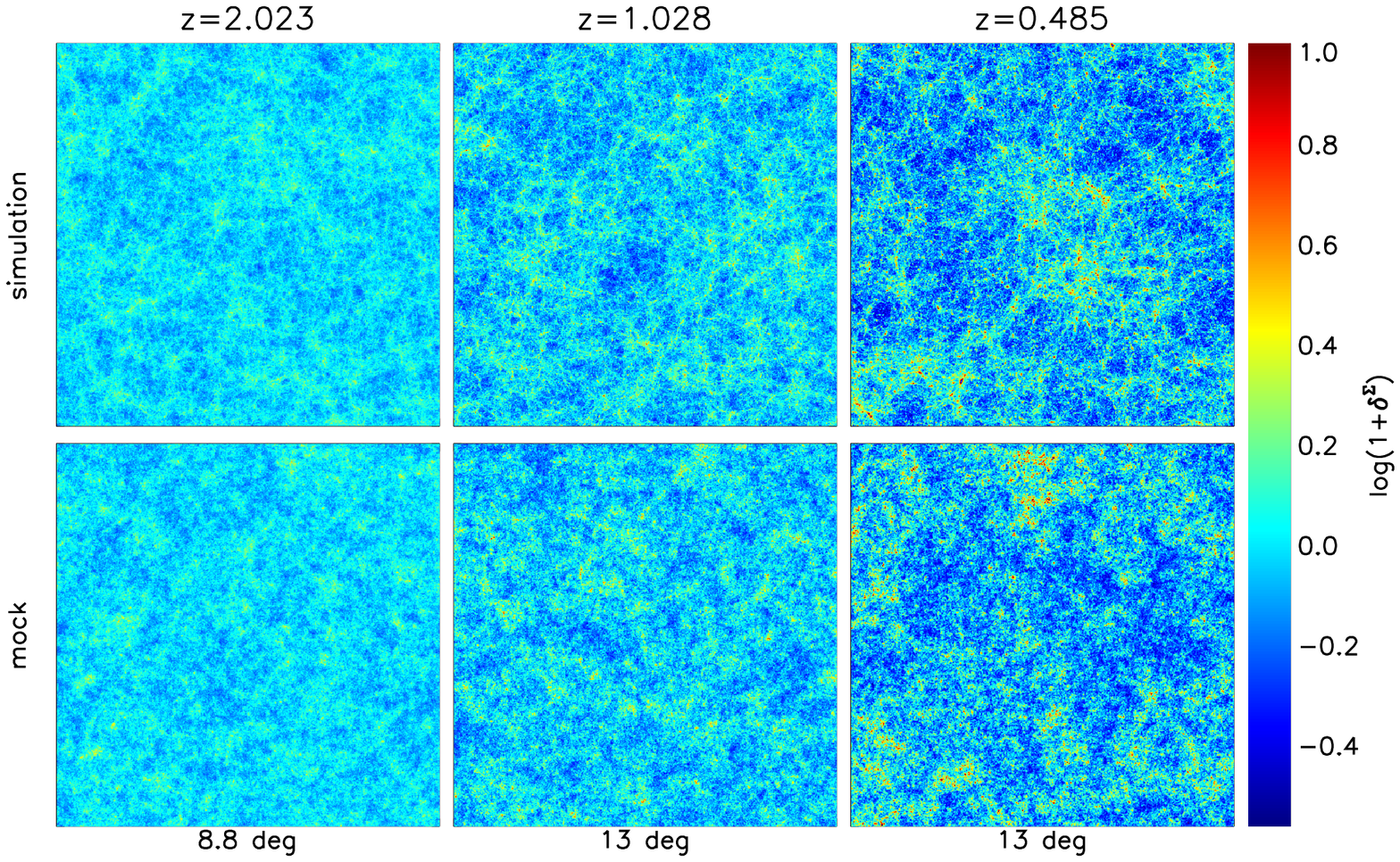}
\caption{The projected density fields from simulation and the inverse-Gaussianization method are presented in the top and bottom rows 
and for $z=2.023$, $1.028$, and $0.485$ from left to right, respectively.
Note that all the panels are different realizations.
The color scale is logarithmic to better show the structures.
The upper and lower plots are quite similar, except that the simulated ones contain many filamentary structures, while the mock ones do not.
This difference is more obvious for lower redshift.}
\label{fig:directview}
\efig

\subsection{Direct view}
\label{sec:directview}

Applying the above local transform, we obtain $48$ maps of $y$ and the
corresponding $y$ power spectra. 
From the averaged $y$ power spectrum (256 linear $\ell$ bins version, different from the plots below),
 we produce 48 Gaussian random $y$ fields with the map size two times
 the original one on a $1024\times1024$ uniform grid. 
Only the center data block on a $512\times512$ uniform grid is taken to
mimic the nonperiodical condition. We then apply the
inverse-Gaussianization with the $y$-$\delta^\Sigma$ relation in
Fig. \ref{fig:transforms} to the above $y$ maps and obtain
realizations of $\delta^\Sigma$ maps. We can then measure various
statistics of the newly generated $\delta^\Sigma$ maps and compare to
the realistic ones directly from the simulation. 

First, we arbitrarily choose one projected density field realization from the simulation and the inverse-Gaussianization method for each redshift and present them together in Fig. \ref{fig:directview}.
The projected density field from simulation and our fast mock generation method look quite similar to the eye.
However, careful readers may find differences between them for $z=0.485$.
The simulated one contains numerous filamentary structures, while the one produced by our fast mock generation method misses them.
The filamentary structures are the result of nonlinear evolution of the cosmic webs.
The information for these structures is largely contained in the three-point correlation function, or the bispectrum (also see Ref. \cite{Obreschkow13} for line correlation as a tracer for filamentary structure.).
Although the Gaussianization could suppress the bispectrum toward zero, we found residual non-Gaussianity left in some configurations in the previous work \cite{yuyu11}.
The inverse-Gaussianization method ignores the above residual non-Gaussianity and thus cannot reproduce well the filamentary structures.
This presents a limitation for the inverse-Gaussianization method.
This limitation is also presented as the failure in reproducing the bispectrum dependence on the configuration presented in the following section.
However, the real weak lensing signal is a weighted summation over several projected density fields, which suppresses the non-Gaussianity to some extent and lowers the importance for the individual filamentary structure.
In that case, the difference would be hard to distinguish like the high redshift case $z=2.023$.

\begin{figure*}
\epsfxsize=8cm
\epsffile{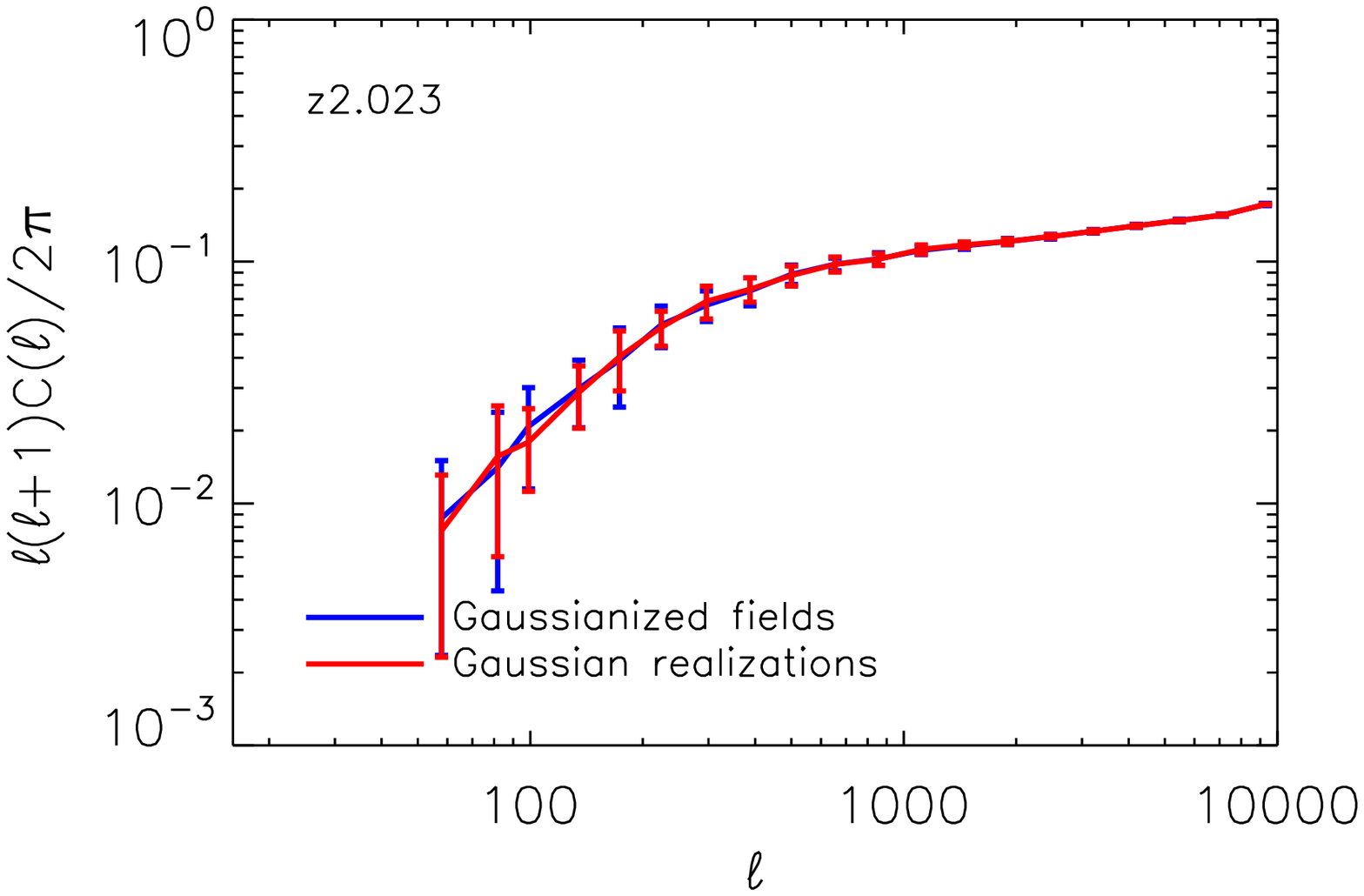}
\epsfxsize=8cm
\epsffile{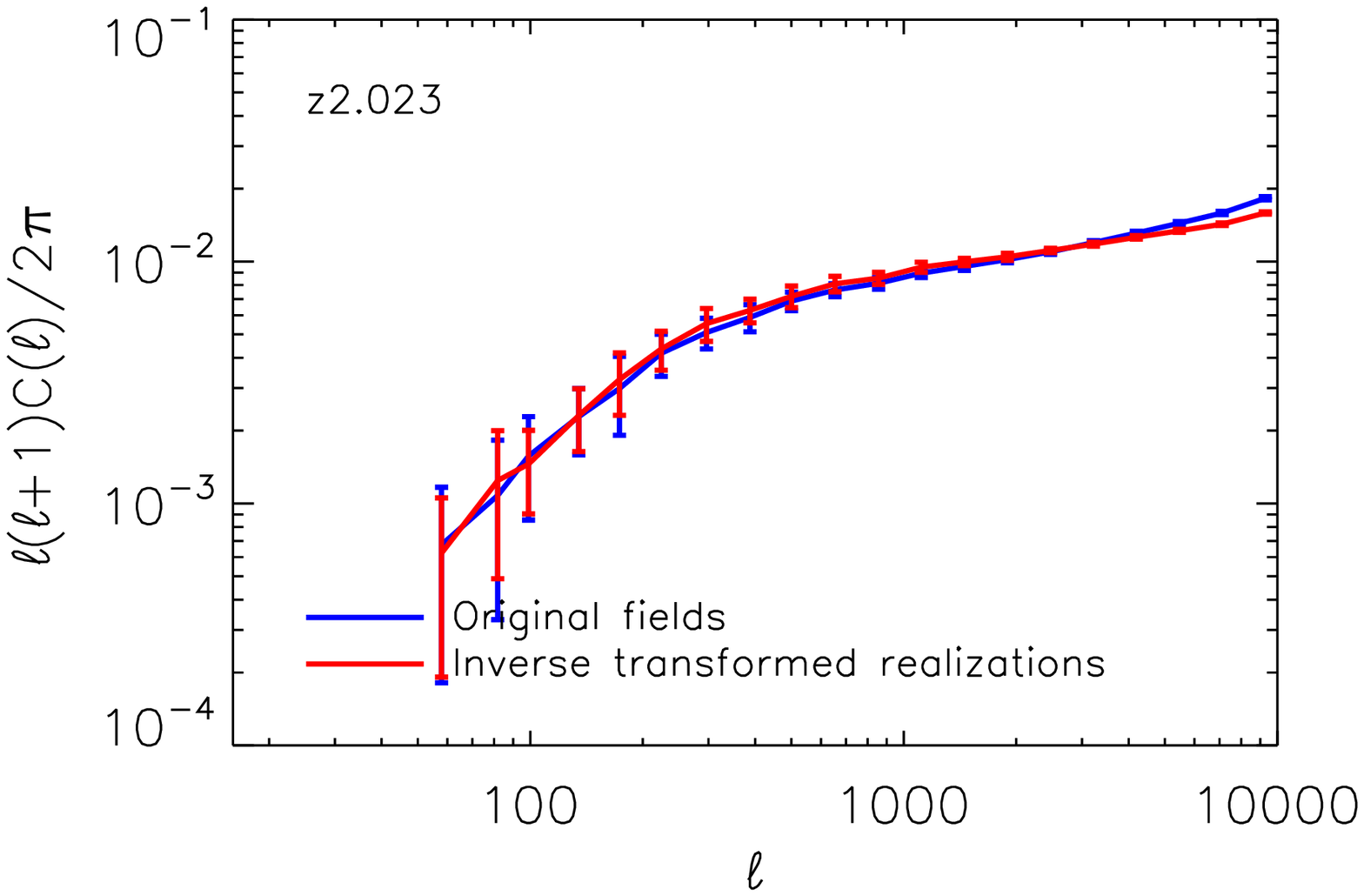}
\\
\epsfxsize=8cm
\epsffile{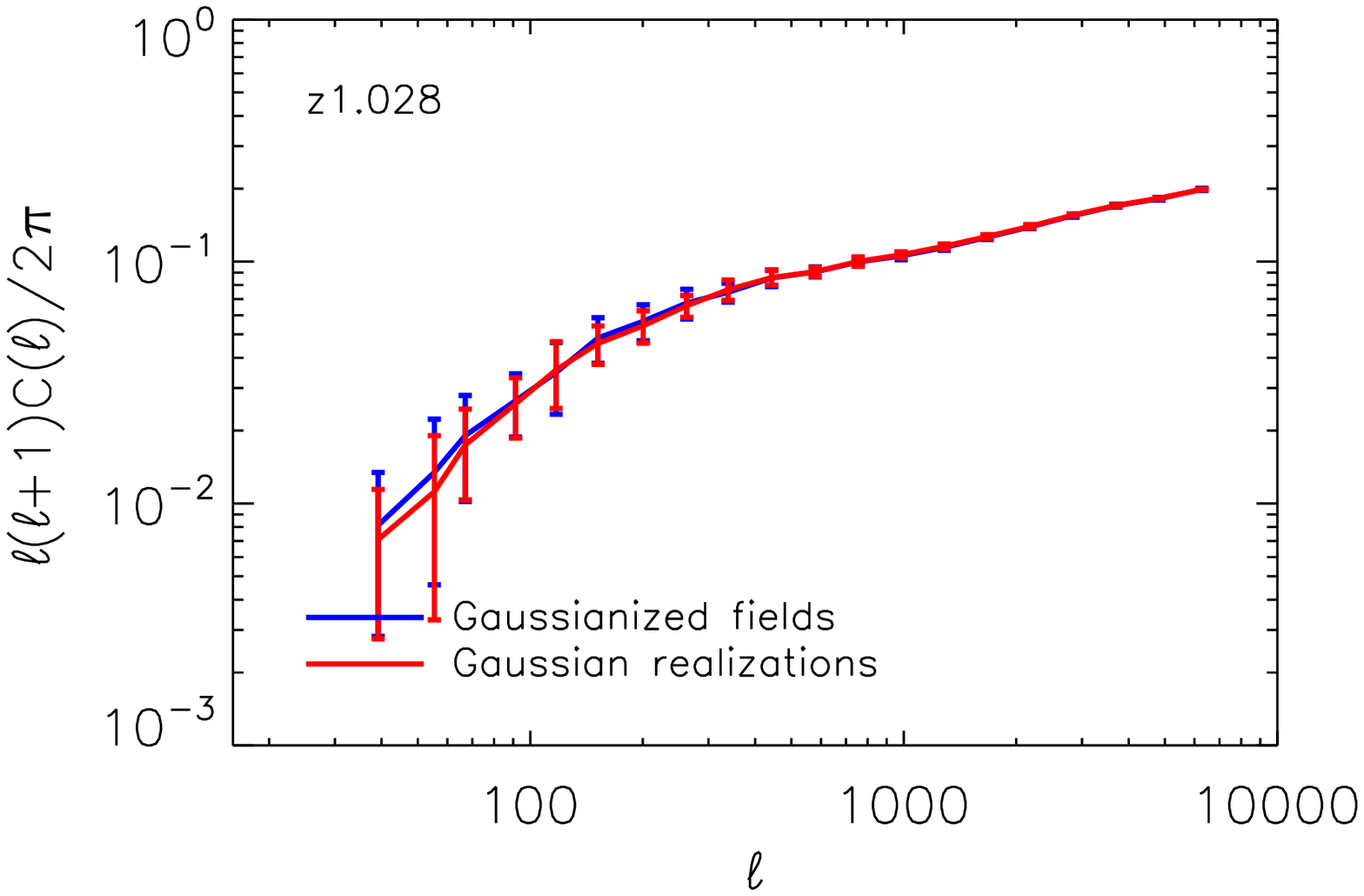}
\epsfxsize=8cm
\epsffile{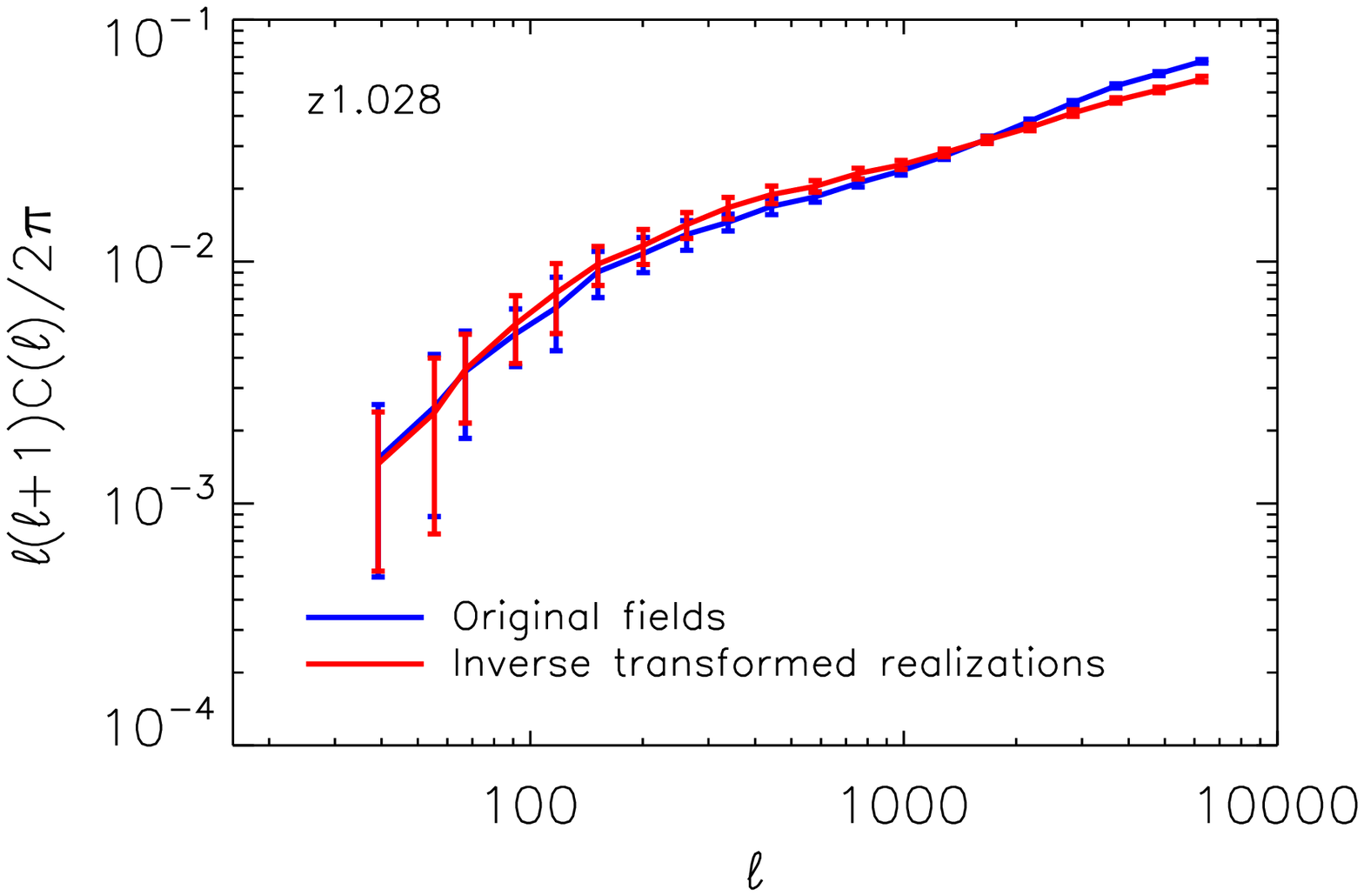}
\\
\epsfxsize=8cm
\epsffile{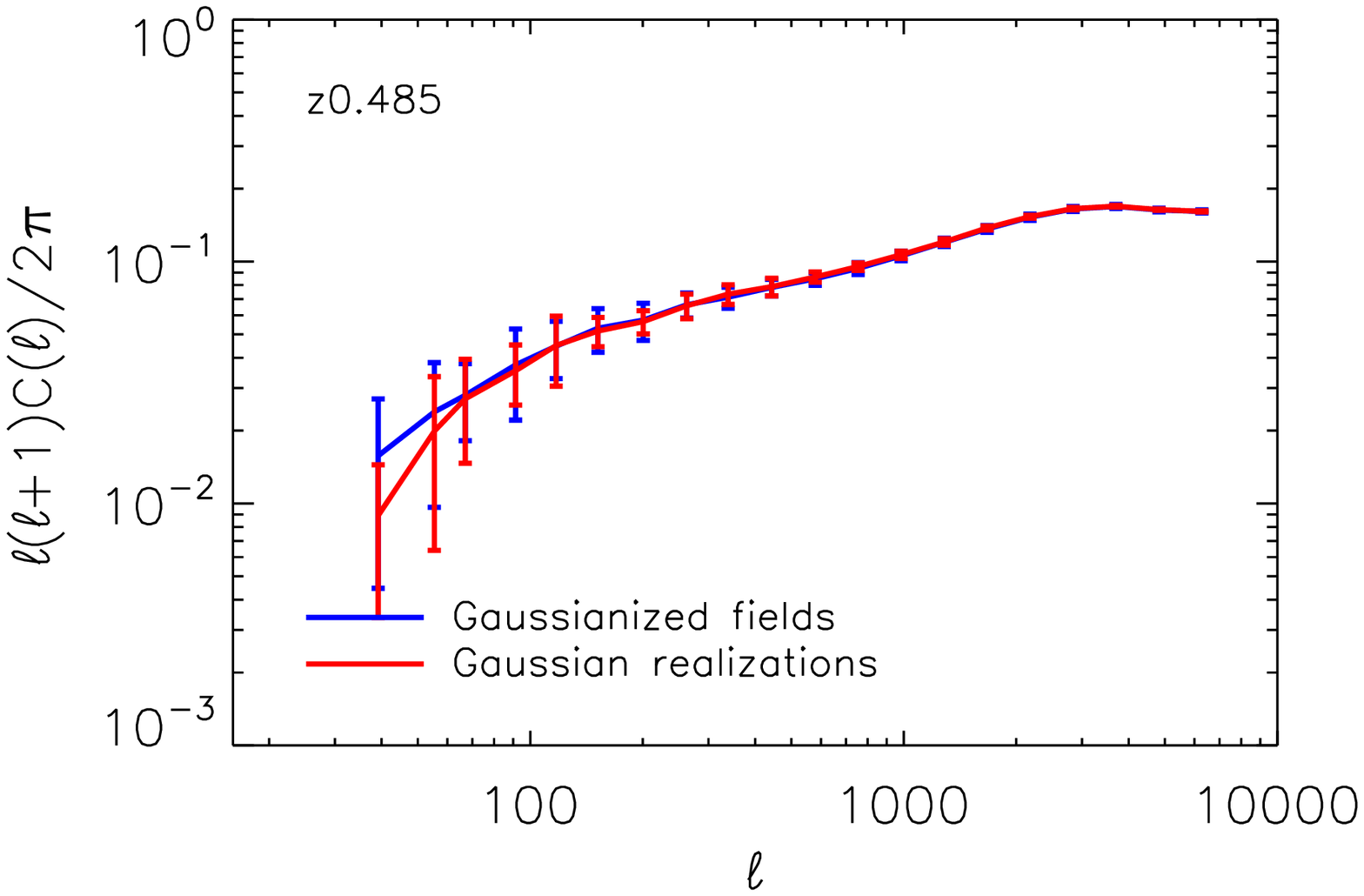}
\epsfxsize=8cm
\epsffile{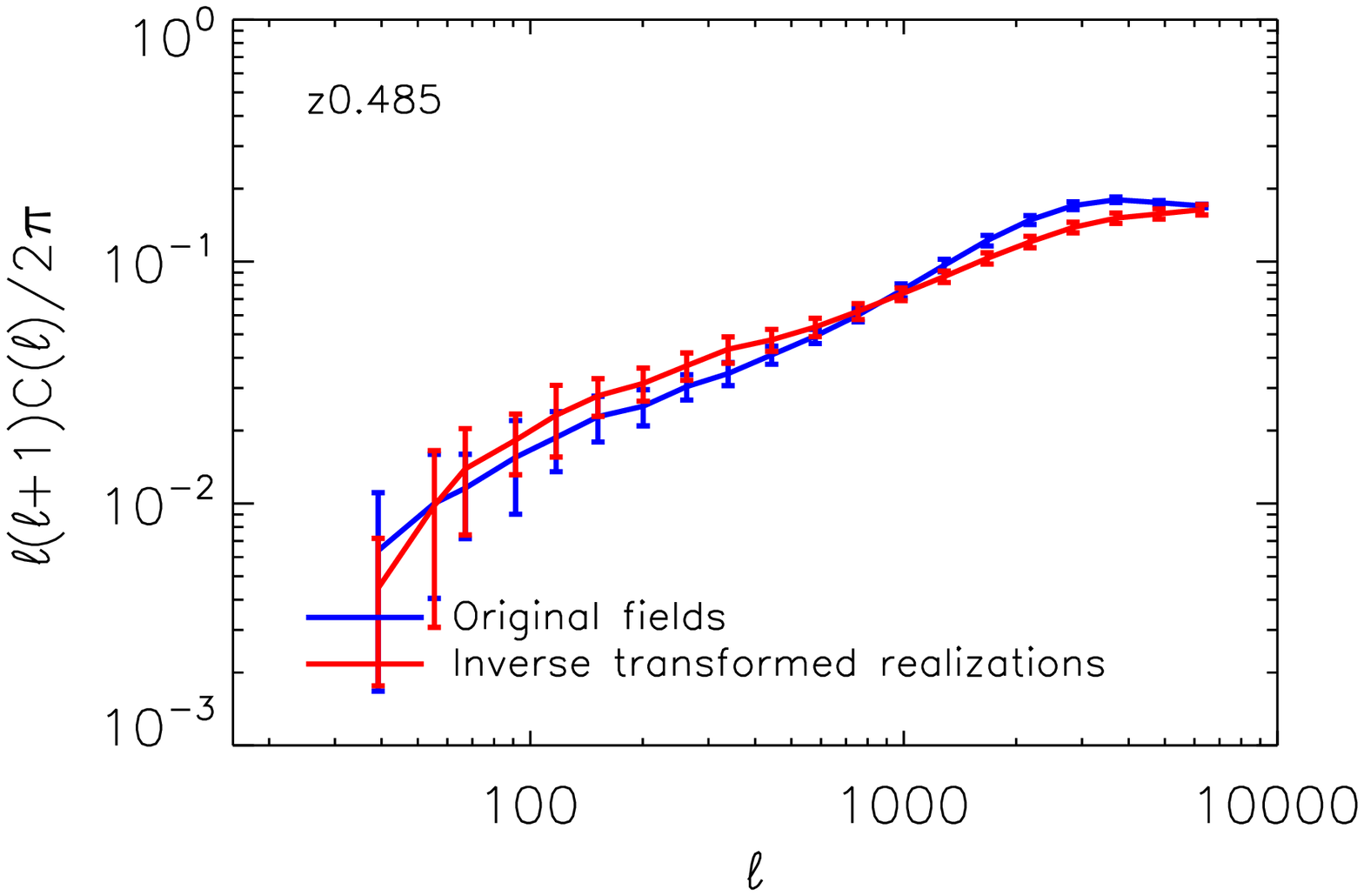}
\caption{Left panels show the power spectra $C_y(\ell)$ of the Gaussianized $y$
  fields  (blue lines) at three redshifts.   From top to bottom, the
  results are for $z=2.023$, $1.028$ and $0.485$, respectively.  As a
  reminder, these redshifts correspond to lens redshifts instead of
  source redshifts.
All the error bars are the r.m.s. among realizations. These results correspond to {\bf
    Step 2} of the inverse-Gaussianization method, detailed in \S
  \ref{sec:method}. {\bf
    Step 3} of the inverse-Gaussianization method generates Gaussian
  Random Fields (GRFs) with $C_y(\ell)$ as the input power
  spectrum. We plot the power spectra of the generated GRFs (red
  lines), just as a trivial test that the generated GRFs indeed process
  identical power spectra as the original $y$ fields.  {\bf
    Step 4} of the inverse-Gaussianization method uses the
  Gaussianization function obtained in {\bf Step 1} (Fig. \ref{fig:transforms}) to
  transform the $y$ maps into $\delta^\Sigma$ maps. Right panels show
  the power spectra of these generated $\delta^\Sigma$ maps (red
  lines). As comparisons, we show the power spectra of the original
  $\delta^\Sigma$ maps (blue lines). At $z=1$ and $z=2$, our method well reproduces the power
  spectra at $\ell\la 2000$ of particular interest to weak lensing
  cosmology. The performance degrades towards lower redshift. At
  $z=0.485$, the power is overproduced at $\ell\la  600$, but
  underproduced at $\ell\ga 1000$. Nevertheless, the error is usually
  smaller than $20\%$ at $\ell\la 10^4$. 
}
\label{fig:powers}
\end{figure*}

\subsection{Power spectra}
\label{sec:power}

We present the power spectra of $y$ and $\delta^\Sigma$ for $z=2.023$, $1.028$, and $0.485$ in the top, middle, and bottom rows of Fig. \ref{fig:powers}.
For each redshift, the power spectrum for the simulation realizations is presented in blue in the right panel.
The power spectrum for the Gaussianized fields (produced in step 2) is presented in blue in the left panel.
The power spectrum for Gaussian random fields produced in step 3 is presented in red in the left panel.
The power spectrum of these mocks (produced in step 4) is presented in red in the right panel.
The error bars are the rms among 48 realizations/mocks.

The left plots are trivial results.
The produced GRFs have the power spectrum consistent with the input one.
The difference may appear for the first several bins due to the cosmic variance.
Nevertheless, considering the error bars they are fully consistent.

From the right plots, we found the whole power spectrum amplitude is produced well.
However, there is difference between simulation realizations and mocks at intermediate and small scales.
At intermediate scales, the power spectrum of mocks is systematically larger than the one of simulations, while at small scales, the mocks have a smaller power spectrum than the simulations.
Thus, the inverse-Gaussianization method is not perfect at reproducing the power spectrum.
The difference is larger for lower redshift.
Thus, we argue that the residual non-Gaussianity in $y$ maps is the reason for this imperfection.
In other words, the Gaussian Copula Hypothesis \cite{Scherrer10} only
holds approximately.
This could also been seen from the residual bispectrum in the
Gaussianized lensing convergence field in Ref. \cite{yuyu11}.  Whether or not we
can take this residual non-Gaussianity into account is an issue for
further investigation. 

Nevertheless, the main purpose of mock is not to produce the accurate
lensing power spectrum. Instead, it is to evaluate the noise in the
power spectrum. Therefore, we proceed to compare the power spectrum
covariance matrix of mocks with that from the simulation. 

\subsection{Power spectrum covariance matrix}

The lensing power spectrum is the most important statistics in weak
lensing cosmology. Therefore, its covariance matrix is of crucial
importance in cosmological parameter constraints. 
The full power spectrum covariance matrix is defined as $\langle
\Delta C(\ell_i)\Delta C(\ell_j)\rangle$. Here, $\Delta C(\ell_i)=C(\ell_i)-\langle C(\ell_i)\rangle$.  Both the diagonal and off-diagonal
elements are essential in cosmological parameter fitting.  The diagonal elements
quantify the error bars of the power spectrum. This is all we may need
to constrain cosmology from the  power spectrum measurement at a
single $\ell$ bin. However,  in reality, we have more than one bin and
errors of the power spectra in these bins can be correlated. These
correlations are described by the off-diagonal
elements. 

\begin{table}
\begin{tabular}{c|c|c|c}
\hline\hline
$N_\ell$ & $r (z=2.023)$ & $r (z=1.028)$ & $r (z=0.485)$ \\
\hline\hline
10 & 1.003 $\pm$ 0.153 & 1.099 $\pm$ 0.205 & 1.315 $\pm$ 0.185\\ 
11 & 1.010 $\pm$ 0.148 & 1.109 $\pm$ 0.198 & 1.291 $\pm$ 0.192\\   
12 & 1.031 $\pm$ 0.158 & 1.121 $\pm$ 0.194 & 1.300 $\pm$ 0.187\\    
13 & 1.025 $\pm$ 0.153 & 1.132 $\pm$ 0.190 & 1.268 $\pm$ 0.212\\  
14 & 1.035 $\pm$ 0.151 & 1.120 $\pm$ 0.188 & 1.232 $\pm$ 0.242\\   
15 & 1.026 $\pm$ 0.150 & 1.110 $\pm$ 0.185 & 1.200 $\pm$ 0.264\\    
16 & 1.023 $\pm$ 0.143 & 1.106 $\pm$ 0.180 & 1.185 $\pm$ 0.262\\   
17 & 1.017 $\pm$ 0.143 & 1.091 $\pm$ 0.185 & 1.175 $\pm$ 0.257\\  
18 & 1.011 $\pm$ 0.141 & 1.093 $\pm$ 0.180 & 1.196 $\pm$ 0.263\\    
19 & 0.993 $\pm$ 0.157 & 1.103 $\pm$ 0.179 & 1.224 $\pm$ 0.283\\
20 & 0.974 $\pm$ 0.174 & 1.118 $\pm$ 0.187 & 1.290 $\pm$ 0.398\\   
\hline\hline
\end{tabular}
\caption{The mean and scatter of the error bar size ratio over the first $N_\ell$ bins.}
\label{tab:compareerror}
\end{table}

\subsubsection{Power spectrum error bars}
\label{sec:error bar}

The rms of the power spectrum on the $i$th $\ell$ bin for the
simulation/mock field is given by the diagonal element of the
covariance matrix, 
\ba
\sigma(C_\mathrm{simu}(\ell_i))=\sqrt{\langle \Delta
  C_\mathrm{simu}(\ell_i)^2\rangle}\ , \no\\
\sigma(C_\mathrm{mock}(\ell_i))=\sqrt{\langle \Delta C_\mathrm{mock}(\ell_i)^2\rangle}\ .
\ea
If there is no correlation among different $\ell$ bins, the sum of the inverse of the band power error bar size is equivalent to the cumulative information content.
Thus, the size of error bars is also a reflection for the cosmological constraint power.
We investigate the mean and scatter of the error bar size ratio over the first $N_\ell$ $\ell$ bins,
\ba
r=\frac{1}{N_\ell}\sum_{i=1}^{N_\ell} r_i=
\frac{1}{N_\ell}\sum_{i=1}^{N_\ell} \frac{\sigma(C_\mathrm{mock}(\ell_i))}{\sigma(C_\mathrm{simu}(\ell_i)}\ ,\\
\sigma^2(r)=\frac{1}{N_\ell}\sum_{i=1}^{N_\ell} (r_i-r)^2\ .
\ea
The result for three redshifts is listed in Table
\ref{tab:compareerror}. 
In weak lensing cosmology, powers at $\ell\lesssim 2000-3000$ are usually adopted in the analysis.
Thus, we care about the averaged ratio of the error bar size for the first $N_\ell\sim 16$ bins.

For high redshift $z=2.023$, the averaged ratio deviates from unity only at $1\%$--$3\%$ level, with the scatter of $\sim 15\%$.
Larger deviation could been seen for lower redshifts.
For $z=1.028$, we find a $\sim 10\%$ larger error bar for the mock power spectrum for all $N_\ell$ we considered.
For the lowest redshift case, the enlargement of the error bar is significant.
The amplitude reaches $\sim 20\%$ at $N_\ell=16$.
Thus, we expect degraded constraint power by using the mock covariance.

\begin{figure}
\epsfxsize=8cm
\epsffile{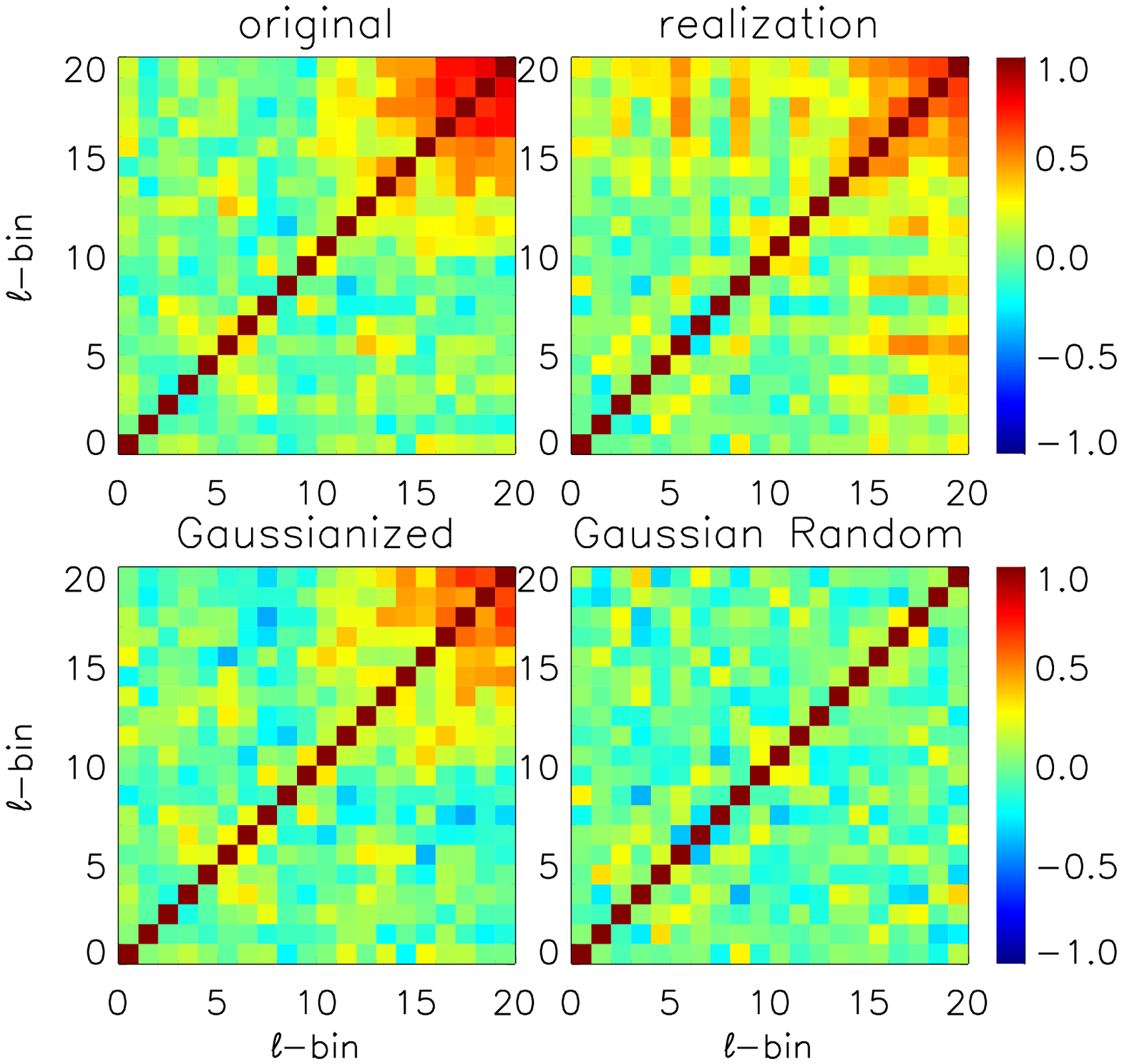}
\caption{Cross-correlation coefficients between power spectra of
  different $\ell$ bins,  for the original $\delta^\Sigma$ fields (top
  left panel), Gaussianized fields ($y$ fields, bottom left), Gaussian random
  $y$ fields (bottom right panel), and the $\delta^\Sigma$ field
  generated by the inverse-Gaussianization method (top-right
  panel). For brevity, readers may only need to compare the top-left
  panel and the top-right panel. It shows that the
  inverse-Gaussianization method basically correctly captures the correlation in
  cosmic variance between power spectra of different $\ell$. This figure is for the
  lens redshift $z=2.023$, highly relevant to the CMB 
  lensing analysis. It is also relevant to deep lensing surveys such
  as LSST and HSC, although to a weaker extent. 
Only the bin label (number) is presented.  For the corresponding $\ell$
value, please refer to Fig. \ref{fig:powers}.} 
\label{fig:cov2-1}
\end{figure}

\begin{figure}
\epsfxsize=8cm
\epsffile{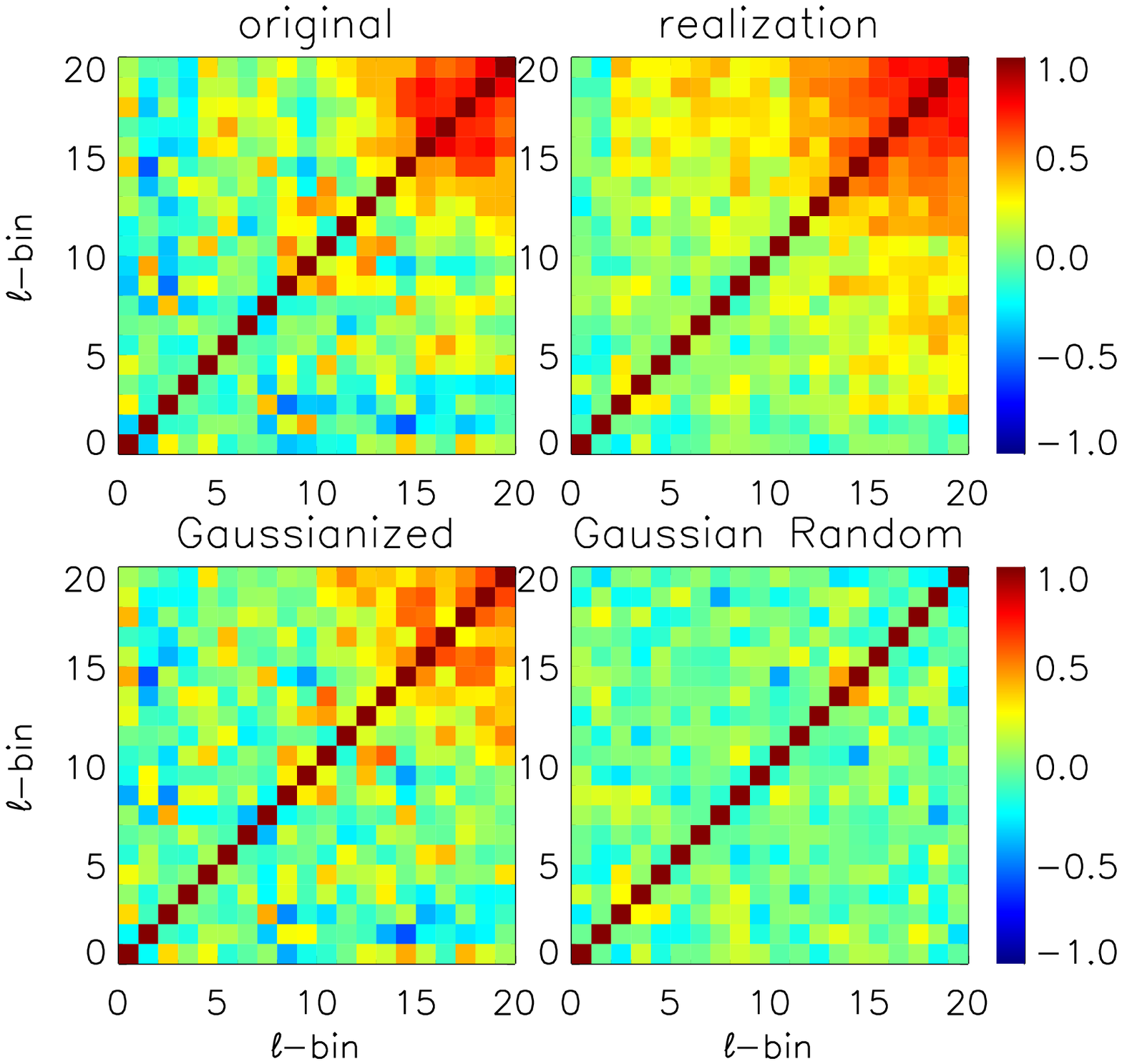}
\caption{Identical to Fig. \ref{fig:cov2-1}, but for the lens redshift
  $z=1.028$. Again, the similarity between the top-left panel (the
  true $\delta^\Sigma$ field) and top
  right panel (the $\delta^\Sigma$ field generated by our method)
  demonstrates the reasonably good performance of our
  method. Nevertheless, we notice off-diagonal elements of the large
  value in the Gaussianized fields (bottom left panel), which are
  statistically significant (compared to the Gaussian random
  realizations, bottom right panel). These significant nonzero
  off-diagonal elements reflect residual non-Gaussianity in the
  Gaussianized $y$ maps and therefore the imperfection of
  Gaussianization. These propagate (nonlinearly) into the final
  $\delta^\Sigma$ maps we generate and cause a statistically significant
difference between the top panels. }
\label{fig:cov2-2}
\end{figure}

\begin{figure}
\epsfxsize=8cm
\epsffile{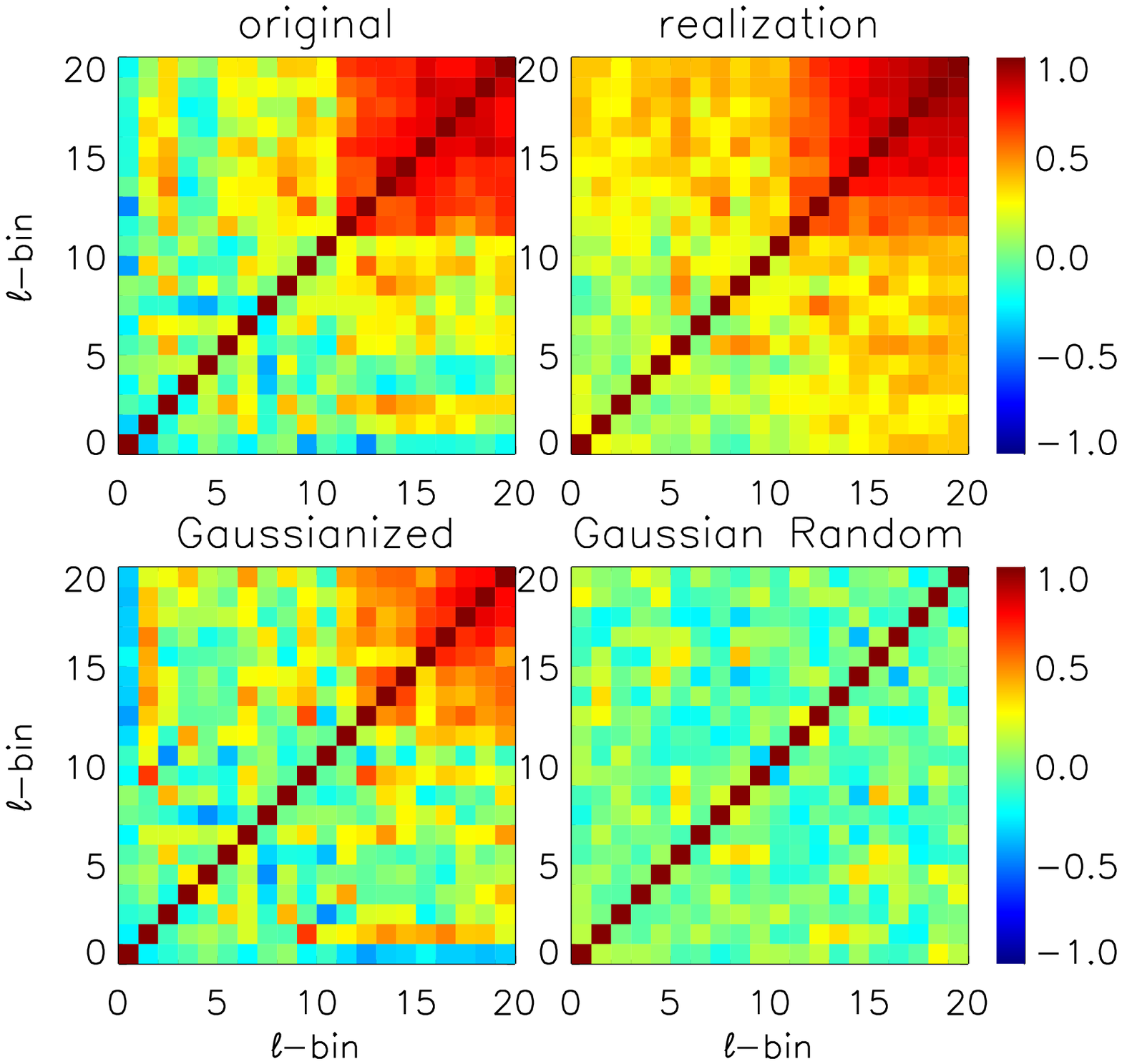}
\caption{Identical to Figs. \ref{fig:cov2-1} and \ref{fig:cov2-2}, but
  for the lens redshift $z=0.485$. Since this is the lowest redshift
  we investigate,  the original $\delta^\Sigma$ map shows the strongest non-Gaussianity and therefore
  the most significant off-diagonal elements (top-left
  panel). Nevertheless, our inverse-Gaussianization method still works
  and well reproduces the correlations between different $\ell$ bins.}
\label{fig:cov2-3}
\end{figure}

\subsubsection{Off-diagonal covariance matrix elements}
\label{sec:correlation}

The off-diagonal elements of the power spectrum covariance matrix are
often presented by the cross-correlation coefficients,
\be
\rho(\ell_i,\ell_j)=\frac{\langle \Delta C(\ell_i) \Delta C(\ell_j) \rangle}
{\sqrt{\langle \Delta C(\ell_i)^2\rangle \langle\Delta C(\ell_j)^2}\rangle}\ .
\ee
The correlation coefficients among $\ell$ bins are presented in Fig. \ref{fig:cov2-1}--\ref{fig:cov2-3} for  $z=2.023$, $1.028$, and $0.485$, respectively.

For all redshifts, we can see strong mode coupling in the simulation realizations (top-left panels).
Small-scale modes not only couple among themselves but also couple with large-scale modes.
The mode coupling is stronger for lower redshift, for the larger nonlinear effects.

If the Gaussian Copula Hypothesis is perfect, we expect a vanishing nondiagonal part in the covariance for Gaussianized fields.
We indeed find that the mode coupling is alleviated in the Gaussianized fields (bottom-left panels) for all redshifts.
The correlation amplitude is suppressed and the area with strong correlation shrinks toward small scales.
However, the correlation is not fully suppressed to zero.
This implies that the Gaussianization method is not perfect.
The correlation between different $\ell$ bins is suppressed but not fully vanished.

The bottom right panels show the trivial results, indicating that 
for the GRFs produced with given power there is no mode coupling by construction.
After applying the inverse transform, mode coupling is indeed recovered at similar scales as the simulated one (top-right panels).
Thus, although the power spectrum and covariance pattern are not
exactly the same as the simulated ones,
these mocks produced by the inverse-Gaussianization method are
reasonably close to realistic ones.

\bfig{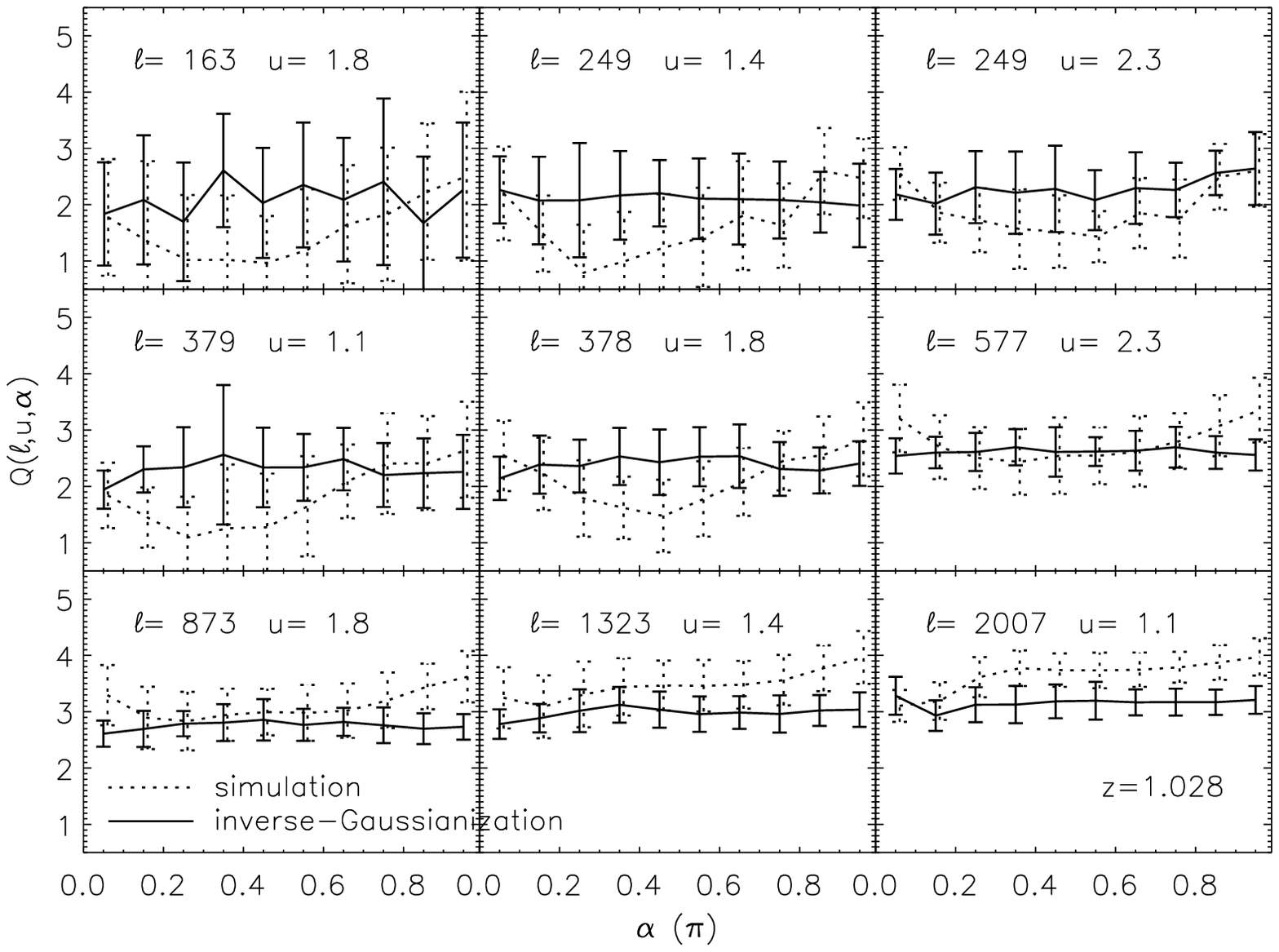}
\caption{The reduced bispectrum $Q(\ell,u,\alpha)$ of the mocks produced by the inverse-Gaussianization method is presented in the solid line for various configurations at $z=1.028$.  For comparison, the reduced bispectrum of simulation slices is presented in the dotted line.
The overall behavior of the reduced bispectrum is reproduced well, except for the even weaker $\alpha$ dependence than the simulated one.
\label{fig:bps}}
\efig

\subsection{Impacts on parameter fitting against the power spectrum}
\label{sec:fit}

Since the error bar size on different $\ell$ bins also correlates,
it cannot be directly translated into the cosmological parameter constraint power.
To compare the difference between adopting the simulated covariance and the mock one produced by the inverse-Gaussianization method,
 we fit the observed power spectrum of each simulation slice to the ensemble one
by a single parameter model, $C(\ell_i)=AC_\model(\ell_i)$.
Here, $A$ is the fitting parameter.
The difference in the statistics of $A$ describes the difference between adopting the covariance measured from the mocks and from the simulations.
As seen in the previous results, the mock fields have different powers from the simulated fields,
leading to the difference in the diagonal part of $\bm{C}_\simu$ and  $\bm{C}_\mock$.
To be fair, we scale the mock covariance  
\be
\bm{C'}_{\mock,ij}= \bm{C}_{\mock,ij}
\frac{C_\simu(\ell_i)}{C_\mock(\ell_i)}
\frac{C_\simu(\ell_j)}{C_\mock(\ell_j)}\ ,
\ee
and drop the prime hereafter.
Since usually in weak lensing we only use the data at $\ell\lesssim 2000-3000$, only the data and covariance in the first $N_\ell=16$ bins are used in the fitting.

We assume that the data points in $\bm{d}=[C(\ell_i), i=1,\cdots, N_\ell]$ follow Gaussian error distribution
 and that the covariance $\bm{C}$ is constant in the parameter space.
$A$ is fit by minimizing
\be
\chi^2(A)=\sum_{i,j=1}^{N_\ell} (d_i-A {d}_{\model,i})[\bm{C}^{-1}]_{ij}(d_j-A {d}_{\model,j})\ .
\ee
Actually, we have analytical expression for the best fit $A$ ,
\be
A=\frac{\bm{d}\bm{C}^{-1}\bm{d}_\model}{\bm{d}_\model\bm{C}^{-1}\bm{d}_\model}\ .
\ee

By definition, $\langle A\rangle=1$ since $\bm{d}_{\model}$ is just the averaged power spectrum.
For $z=2.023$, among 48 fittings, $\sigma(A)=0.013$ and $0.020$ is observed for using $\bm{C}_\mathrm{simu}$ and $\bm{C}_\mathrm{mock}$, while for $z=1.028$, $\sigma(A)=0.015$ and $0.020$ for adopting $\bm{C}_\mathrm{simu}$ and $\bm{C}_\mathrm{mock}$. 
For $z=0.485$, we obtain $\sigma(A)=0.039$ and $0.068$ for using $\bm{C}_\mathrm{simu}$ and $\bm{C}_\mathrm{mock}$.
This result is consistent with the error bar size analysis in the previous section.
Using mock covariance leads to degradation in the parameter
constraint.   Whether we can produce better mocks relies on whether we
can improve the Gaussianization and the corresponding
inverse Gaussianization. This is an issue for further investigation. 

\subsection{Lensing bispectrum}
\label{subsec:bispectrum}

Another important weak lensing statistics is its bispectrum. It is
highly complementary to the lensing power spectrum. The lensing power
spectrum suffers from a degeneracy between $\Omega_m$ and
$\sigma_8$. However, this can be broken by the lensing bispectrum (or
skewness), which is mainly sensitive to $\Omega_m$. The lensing
skewness has been detected for more than a decade (e.g. Refs. \cite{Pen03,Jarvis04}). The
three-point correlation function and bispectrum in general have been
robustly measured by CFHTLenS \cite{fulp14}. 

Our inverse-Gaussianization method is also capable of reproducing the
lensing bispectrum. For one example, we show the comparison of reduced
bispectrum $Q(\ell,u,\alpha)$ in Fig. \ref{fig:bps}. The reduced
bispectrum is defined as 
\be
Q(\vec\ell_1,\vec\ell_2,\vec\ell_3)=
\frac{B(\vec\ell_1,\vec\ell_2,\vec\ell_3)}
{C_1C_2+C_2C_3+C_3C_1}\ .
\ee
Here, $C_i\equiv C(\ell_i)$. 
The result is presented in the form of $Q(\ell,u,\alpha)$, in which
$\vec\ell_1$ is the one with the smallest amplitude among
$(\vec\ell_1,\vec\ell_2,\vec\ell_3)$, $\ell=|\vec\ell_1|$,
$u=\ell_2/\ell_1$ and $\alpha$ is the angle between $\vec\ell_1$ and
$\vec\ell_2$.  Unlike the reduced bispectrum of
  3D density, which shows significant dependence on $\alpha$ (e.g., Ref. \cite{guohong09}), the 2D
  reduced bispectrum in simulation has only weak dependence on
  $\alpha$. But the $\alpha$ dependence in maps generated by our
  method is likely even weaker.  
  This result is consistent with the lack of filamentary structures in the produced mocks in Fig. \ref{fig:directview}.
  This is likely a generic
  consequence of the inverse-Gaussianization method. In the Appendix
  we show an argument using lognormal fields as an example.
  
Nevertheless, our method reproduces the overall behavior of the reduced
  bispectrum over all the configurations investigated. This makes our
  method useful to generate lensing mocks for lensing bispectrum
  analysis.  Since our method reproduces the lensing power
  spectrum, bispectrum and trispectrum
  (power spectrum covariance) reasonably well, we expect that lensing
  statistics such as peak abundance \cite{fanzh10,liuxk2015,liujia15} may also be reproduced well. These
statistics will be studied elsewhere. 

\begin{figure*}
\epsfxsize=8cm
\epsffile{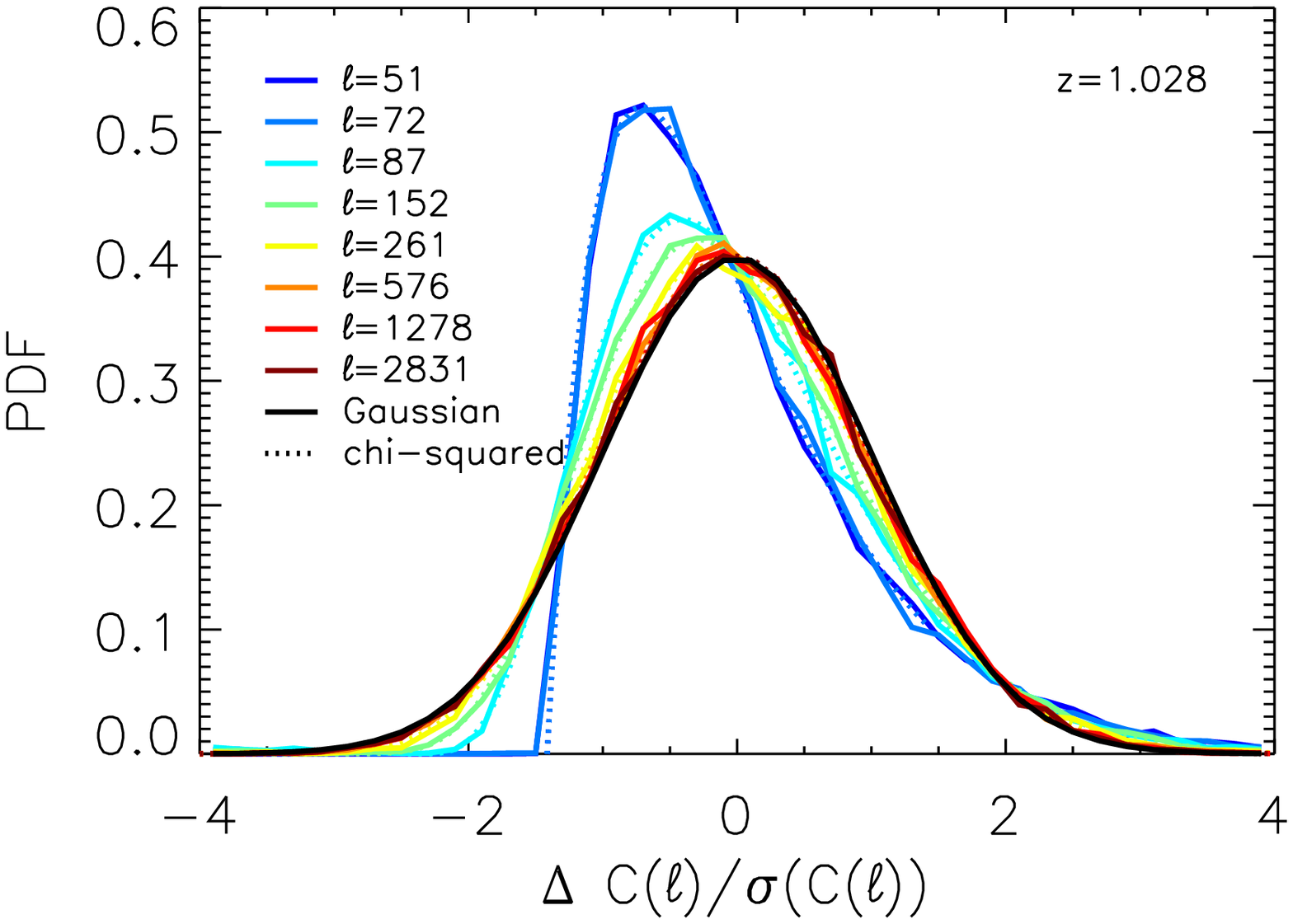}
\epsfxsize=8cm
\epsffile{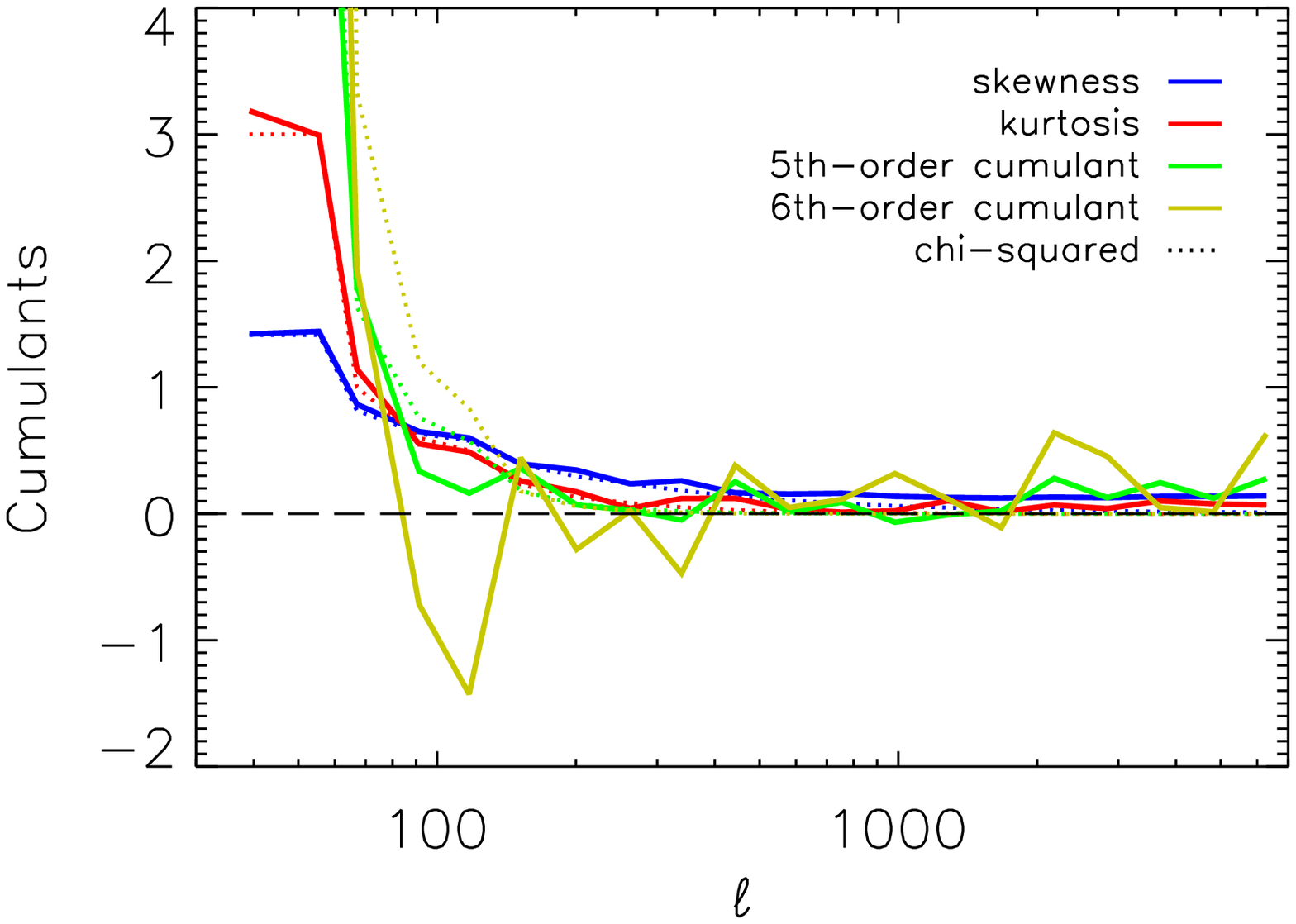}
\caption{The solid lines in the left panel present the PDF of the band power at eight arbitrarily selected scales for $z=1.028$.
The PDF is measured from 16384 mocks, and the bin size for the $x$ axis is $0.2$.
The corresponding $\chi^2$-distribution is plotted in the dotted line, while the Gaussian distribution is the black solid line.
Increasing non-Gaussianity is observed toward decreasing $\ell$, 
but it is quite consistent with the $\chi^2$-distribution.
The right panel presents the cumulants of band power distribution up to sixth order,
 along with the cumulants of the corresponding $\chi^2$-distribution.
Toward large $\ell$, the band power cumulants reduce toward zero.
However, residual skewness is still observed.
\label{fig:pdf}}
\end{figure*}


\section{Application}
\label{sec:application}

In the single parameter fitting analysis in the previous section,
we assume that the power spectrum error distributes as Gaussian.
This is also a widely used assumption in the cosmological parameter constraints.
However, theoretically we know that this assumption breaks down at least at large scales.
During the linear evolution from the Gaussian initial conditions, the matter power spectrum at a given wave number $k$ is distributed as $\chi^2$ with the degrees of freedom of $N_k$, the number of $k$-modes in the survey volume.
This is also well studied in the CMB field \cite{Bond98,Verde03}.

In the large $N_k$ limit, which corresponds to sufficiently large volumes and high wave numbers, the above Gaussian assumption is considered as good, due to the central-limit theorem.
However, to which scale the Gaussian limit is reached is not specified.
Furthermore, at high $k$, the nonlinear evolution of matter clustering is expected to introduce non-Gaussianities, leading to the departures from a $\chi^2$ distribution in the large $N_k$ limit.

The upcoming generation of surveys will probe the distribution of matter in the Universe with unprecedented precision.
Accurate parameter constraints not only require accurate covariance and precision matrix estimation
 but also require the prior knowledge of the joint band power probability distribution.
If the non-Gaussian feature in the band power distribution is not correctly concerned, it also leads to bias in cosmological parameters.
However, the study of band power distribution is a challenging work.
It requires a huge amount of realizations.
For example, \citet{Blot15} found the deviation from $\chi^2$ at high $k$ while previous work \cite{Takahashi09} did not, and they concluded that the insufficient number of ensembles used to determine the power spectrum distribution is the reason.

As an example for application, we try to measure the lensing power
spectrum PDF of each $\ell$ sbin from a large amount of mocks produced by the inverse-Gaussianization method.
We arbitrarily produce 16384 mocks for $z=1.028$.
This large number of realizations enables us to obtain good statistics
on an individual $\ell$ bin. If needed, we can also generate many more
mocks to measure the joint PDF of several to many $\ell$ bins. 

The measured band power probability distribution $\mathrm{PDF}(\Delta C(\ell)/\sigma(C(\ell)))$ 
 on eight arbitrarily chosen $\ell$ bins is presented in the left panel of Fig. \ref{fig:pdf},
 along with the corresponding $\chi^2$-distribution with the degrees of freedom of $N_\ell$.
The eight chosen $\ell$ bins range from $\ell=51$ to $2831$, covering the linear to nonlinear regime.
From the plot, significant deviation from the Gaussian distribution is observed for five low $\ell$ bins.
However, they are predicted well by the corresponding $\chi^2$-distribution.
For high $\ell$ bins with $\ell\gtrsim 300$, the resulting distributions are close to Gaussian. 
This is expected from the central-limit theorem for the fact that a large amount of modes contributed to the high $\ell$ bins.

We also calculate up to sixth-order cumulants of the band power distribution as a function of $\ell$.
The result is presented in the right panel of Fig. \ref{fig:pdf} in a solid line,
along with the up to sixth-order cumulants for the corresponding $\chi^2$-distribution in a dotted line.
Nonzero band power cumulants are observed at low $\ell$, indicating the non-Gaussianity in the statistics.
Again, at low $\ell$, they are predicted well by the $\chi^2$-distribution except for the sixth-order cumulants which suffer a large statistical error. 
Toward high $\ell$, all the cumulants reduce toward zero.
However, at small scales, the skewness deviates from $\chi^2$ prediction at $\ell\sim 500$ and still shows deviation from zero at even larger $\ell$.
This result is consistent with the result from a large suite of simulations \cite{Blot15}.

The above result has important observational implications that warrant further investigation.


\section{conclusion}
\label{sec:conclusion}

In this work, we proposed a new fast mock generation method, named the
inverse-Gaussianization method. 
From a single N-body/hydro simulation and several realizations of
lensing maps produced from such a simulation, the inverse-Gaussianization
method enables us to quickly produce in principle {\it infinite
  statistically independent} lensing convergence maps, 
almost as fast as producing the 2D simulation initial conditions.
Basically, it has four steps: 
(1) obtaining the local transform to Gaussianize the simulated fields,
(2) applying the averaged local transform to obtain Gaussianized fields,
(3) producing GRFs according to the power spectrum of Gaussianized fields, and
(4) applying the inverse transform on these GRFs to produce realistic mocks.
This is a deterministic procedure since there are no free parameters
involved in the process.

We tested the inverse-Gaussianization method on the lensing
convergence tomography maps generated by matter distribution of $300$ Mpc/$h$
projection length centered at lens redshift $z=2.023$, $1.028$, and
$0.485$. This is a test more stringent than on the lensing convergence
maps along the full light cone to the sources. 
We produced 48 lensing convergence mocks by the inverse-Gaussianization method from the local transform and power spectrum measured from 48 simulated lensing convergence slices.
We compared several statistics, such as the power spectrum, its error bar size, the cross correlation coefficient, the result of single parameter fitting, and the bispectrum.
First, we found the inverse-Gaussianization method reproduces the mode coupling between different scales with a similar pattern as the one of simulated fields.
Second, we found the inverse-Gaussianization method is not perfect but performs reasonably well.
Mild deviations in the power spectrum between the simulated fields and mock fields is observed.
The mock fields produced by the inverse-Gaussianization method have larger power at intermediate scales than the simulated fields and vice versa in the small scales.
The deviation is larger for lower redshifts.
The imperfection of the Gaussian Copula Hypothesis causes these deviations, which could be seen from the residual covariance for the Gaussianized fields.
Third, the power spectrum of 48 mock lensing convergence fields has a slightly larger rms than that of the simulated slices.
The averaged increasing is $\sim 2\%$, $10\%$, and $20\%$ for $z=2.023$, $1.028$, and $0.485$, when only the first 16 bins are considered.
Also, by a simple single parameter fitting test, 
the rms of the fitting parameter $A$ is larger for adopting the covariance obtained from the mock fields.
The increase in $\sigma(A)$ is $0.007$, $0.005$, and $0.029$ for $z=2.023$, $1.028$, and $0.485$, respectively.
Both the increase of error bar size and the increase of the rms in the single fitting parameter $A$ indicate
the degradation of parameter constraint power when the mock covariance is adopted.
We also found a limitation for the inverse-Gaussianization method.
It could not reproduce the filamentary structure well for a low redshift projected density field,
 which is also related to the lack of bispectrum dependence on $\alpha$, the angle between two wave vectors.

Nevertheless, the produced mocks have a realistic and reasonably good power spectrum and covariance.
Observational effects, such as noise, mask, and geometry, could been easily included in these mock fields.
Specifically, the shape noise is very important for realistic weak lensing mocks and is also a dominant source of the statistical errors.
Assuming the shape noise in a Gaussian form, one can easily add this noise into the produced noiseless mocks.
One can also adopt the method proposed by Ref. \cite{Shirasaki16arxiv}, which populates a real catalog of source galaxies into the simulated light cone, erases the original weak lensing signal by random rotation, and adds the mock lensing signal back in.
This method maintains the observed distribution of source galaxies on the sky, their shapes as well as their redshift distribution, and thus more realistic shape noise effects.
We conclude that these mocks could be used in a cosmology study, such as testing statistical tools, quantifying mask and geometry effects, identifying systematics, etc. 
Finally, we measured the probability distribution of band power on several bins from 16384 mock fields produced by the inverse-Gaussianization method as an example of the application.
Significant and not fully vanished non-Gaussianity observed at large and small scales implies the breakdown of the Gaussian assumption for the band power error distribution.

For the presented various comparisons between mocks and simulation realizaitons, the difference is larger for lower redshift.
We argued that the imperfection of Gaussianization leads to these differences.
Note that in this work the simulated lensing convergence is only a projected matter density over $300\mpch$ along the line of sight.
For a more realistic lensing convergence field in the weak lensing survey, it is a projection of matter over a much larger range along the line of sight.
In this case, the non-Gaussianity in the real fields is suppressed due to the projection.
Thus, we expect that for more practical application, the performance is better, as the results for high redshift.


\section{Discussion}
\label{sec:discussion}

Recently, super-sample covariance was found to be a dominant sampling error for matter power spectrum estimators in a finite volume, which arises from the presence of super-survey modes \cite{Takada13}.
The mocks produced by the proposed inverse-Gaussianization method contain a part of the super-sample covariance, since the mocks are cut out from maps two times larger.
One could also include super-sample variance in the mock fields by setting the power at scales larger than the size of the target field.
Therefore, our method is useful for studying the impact of super-sample covariance.
On the other hand, 
a separate universe technique was proposed and was found to be very useful and straightforward for calibrating these sampling errors \cite{liyin14}.
Furthermore, the variance of super-survey modes can be easily computed from Gaussian
realizations of the fields as long as super-survey modes are in the linear regime, which is the case for a wide-area survey.
Reference \cite{Mohammed14} models the super-sample variance effect as an additional parameter that can be determined from the data. These studies greatly reduce the number of realizations needed to calibrate the covariance.

Besides the assumptions widely used for power spectrum covariance, there are several other sources of uncertainty, and bias should be noticed.

In matter clustering analysis, the covariance is most commonly assumed to be constant in parameter space. 
However, \citet{Eifler09} found that cosmic shear covariances vary significantly within the considered parameter range (also see Ref. \cite{Reischke16}). 
The cosmology assumed in the covariance has a non-negligible impact on the size of the likelihood contours. This impact increases with increasing survey size, increasing number density of source galaxies, decreasing ellipticity noise, and when taking non-Gaussianity into account. 
They concluded that a proper treatment of this effect is therefore even more important for future surveys.
In this aspect, fast mock generation methods could help to test this problem,
since a large number of realizations is required for each parameter configuration.

The derived cosmological {\it parameter covariance} also suffers bias and errors from the effects in sampling and modelling the data covariance.
\citet{Taylor14} found that, for sampled data covariances with Gaussian-distributed data,
 the parameter covariance matrix estimated from the width of the likelihood has a Wishart distribution.
Thus, the parameter mean and covariance obtained from this likelihood are biased.
They proposed an unbiased estimator of the parameter covariance matrix as a remedy.
However, the data themselves are not Gaussian distributed.
As stated in Sec. \ref{sec:application}, the band power distribution at both large and small scales deviates from the Gaussian distribution.
To take full advantage of the future high precision surveys, the knowledge on the band power distribution is worth investigation.
In these studies, one could make use the fast mock generation methods.


\section*{Acknowledgment}
This work was supported by the National Science Foundation of China (Grants No. 11403071, No. 11533006 and No. 11320101002), National Basic Research Program of China (973 Programs No. 2015CB857001 and No. 2015CB857003), and the Strategic Priority Research Program ``The Emergence of Cosmological Structures'' of the Chinese Academy of Sciences (Grant No. XDB09000000).
This work made use of the High Performance Computing Resource in the Core
Facility for Advanced Research Computing at Shanghai Astronomical Observatory.

\appendix
\numberwithin{equation}{section}

\section{On the bispectrum of the generated weak lensing fields}
The reduced bispectrum of our generated lensing maps has weaker
dependence on $\alpha$. We speculate it as a generic consequence of
the inverse-Gaussianization, insensitive to details of the
Gaussianization function. To better understand this behavior,  we present a rather heuristic investigation of the bispectrum of a lognormal field. 
A field $\kappa$ being lognormal means that $\kappa=A(e^{y}-\exp(\sigma^2_y/2))$, with $y$ a Gaussian field with $\langle y\rangle=0$, $\sigma^2_y\equiv \langle y^2\rangle$ and $A$ a constant.   Using the cumulant expansion theorem,  the two-point correlation function is 
\be
\xi_{12}(\theta_{12})\equiv \langle \kappa_1\kappa_2\rangle=B^2(e^{\langle y_1y_2\rangle}-1)\ .
\ee
Here $\kappa_i\equiv \kappa(\vec{\theta}_i)$, $y_i\equiv y(\vec{\theta}_i)$, $\theta_{ij}\equiv |\vec{\theta}_i-\vec{\theta}_j|$, and $B\equiv A\exp(\sigma^2_y/2)$. 
The three-point correlation function is
\ba
\zeta_{123}&\equiv\langle \kappa_1\kappa_2\kappa_3\rangle\\
&=& B^3(e^{\langle y_1y_2\rangle+\langle y_2y_3\rangle+\langle y_3y_1\rangle}\no\\
&& -(e^{\langle y_1y_2\rangle}+e^{\langle y_2y_3\rangle}+e^{\langle y_3y_1\rangle})+2)\no\ .
\ea
The three-point (and all higher-point) correlation function(s) is (are) determined by the two-point correlation function, 
\ba
\zeta_{123}&=& B^{-1}(\xi_{12}\xi_{23}+\xi_{23}\xi_{31}+\xi_{31}\xi_{12})\\
&&+B^{-3}\xi_{12}\xi_{23}\xi_{31}\ . \no
\ea
The bispectrum is then
\ba
B_3(\vec{\ell}_1,\vec{\ell}_2,\vec{\ell}_3)&=&\frac{C_1C_2+C_2C_3+C_3C_1}{B}\\
&&+B^{-3}\int \frac{d^2\ell^{'}}{(2\pi)^2}C(\ell^{'})C(\vec{\ell}_2+\vec{\ell}^{'})C(-\vec{\ell}_1+\vec{\ell}^{'})\ . \no
\ea
Here $C_i\equiv C(\ell_i)$ and $C(\ell)$ is the angular power spectrum. Since $\vec{\ell}_1+\vec{\ell}_2+\vec{\ell}_3=0$, the last integral above is actually symmetric between $\vec{\ell}_{1,2,3}$. 

Under the limit $|B^{-2}\xi|\ll 1$, we can neglect the last integral, and the reduced bispectrum is
\be
Q(\vec{\ell}_1,\vec{\ell}_2,\vec{\ell}_3)\simeq B^{-1}\ .
\ee
Therefore, it does not depend on the angles between $\vec{\ell}_{i}$,
nor $\ell_i$. This is actually what we have found in
Fig. \ref{fig:bps}. 

The actual $Q(\vec{\ell}_1,\vec{\ell}_2,\vec{\ell}_3)$ does show
dependences on $\ell$, $u$, and $\alpha$, although weak. Such
dependences may arise from either the deviation from a log-normal
distribution or the neglected $\xi^3$ term. However, we suspect that
the major cause is likely the nonzero off-diagonal
elements shown in Figs. \ref{fig:cov2-1}, \ref{fig:cov2-2}, and \ref{fig:cov2-3}.

\bibliographystyle{apsrev4-1}
\bibliography{mybib150821}

\begin{thebibliography}{67}%
\makeatletter
\providecommand \@ifxundefined [1]{%
 \@ifx{#1\undefined}
}%
\providecommand \@ifnum [1]{%
 \ifnum #1\expandafter \@firstoftwo
 \else \expandafter \@secondoftwo
 \fi
}%
\providecommand \@ifx [1]{%
 \ifx #1\expandafter \@firstoftwo
 \else \expandafter \@secondoftwo
 \fi
}%
\providecommand \natexlab [1]{#1}%
\providecommand \enquote  [1]{``#1''}%
\providecommand \bibnamefont  [1]{#1}%
\providecommand \bibfnamefont [1]{#1}%
\providecommand \citenamefont [1]{#1}%
\providecommand \href@noop [0]{\@secondoftwo}%
\providecommand \href [0]{\begingroup \@sanitize@url \@href}%
\providecommand \@href[1]{\@@startlink{#1}\@@href}%
\providecommand \@@href[1]{\endgroup#1\@@endlink}%
\providecommand \@sanitize@url [0]{\catcode `\\12\catcode `\$12\catcode
  `\&12\catcode `\#12\catcode `\^12\catcode `\_12\catcode `\%12\relax}%
\providecommand \@@startlink[1]{}%
\providecommand \@@endlink[0]{}%
\providecommand \url  [0]{\begingroup\@sanitize@url \@url }%
\providecommand \@url [1]{\endgroup\@href {#1}{\urlprefix }}%
\providecommand \urlprefix  [0]{URL }%
\providecommand \Eprint [0]{\href }%
\providecommand \doibase [0]{http://dx.doi.org/}%
\providecommand \selectlanguage [0]{\@gobble}%
\providecommand \bibinfo  [0]{\@secondoftwo}%
\providecommand \bibfield  [0]{\@secondoftwo}%
\providecommand \translation [1]{[#1]}%
\providecommand \BibitemOpen [0]{}%
\providecommand \bibitemStop [0]{}%
\providecommand \bibitemNoStop [0]{.\EOS\space}%
\providecommand \EOS [0]{\spacefactor3000\relax}%
\providecommand \BibitemShut  [1]{\csname bibitem#1\endcsname}%
\let\auto@bib@innerbib\@empty
\bibitem [{\citenamefont {{Yu}}\ \emph {et~al.}(2011)\citenamefont {{Yu}},
  \citenamefont {{Zhang}}, \citenamefont {{Lin}}, \citenamefont {{Cui}},\ and\
  \citenamefont {{Fry}}}]{yuyu11}%
  \BibitemOpen
  \bibfield  {author} {\bibinfo {author} {\bibfnamefont {Y.}~\bibnamefont
  {{Yu}}}, \bibinfo {author} {\bibfnamefont {P.}~\bibnamefont {{Zhang}}},
  \bibinfo {author} {\bibfnamefont {W.}~\bibnamefont {{Lin}}}, \bibinfo
  {author} {\bibfnamefont {W.}~\bibnamefont {{Cui}}}, \ and\ \bibinfo {author}
  {\bibfnamefont {J.~N.}\ \bibnamefont {{Fry}}},\ }\href {\doibase
  10.1103/PhysRevD.84.023523} {\bibfield  {journal} {\bibinfo  {journal}
  {\bibprd}\ }\textbf {\bibinfo {volume} {84}},\ \bibinfo {eid} {023523}
  (\bibinfo {year} {2011})},\ \Eprint {http://arxiv.org/abs/1103.2858}
  {arXiv:1103.2858 [astro-ph.CO]} \BibitemShut {NoStop}%
\bibitem [{\citenamefont {{Scoccimarro}}\ \emph {et~al.}(1999)\citenamefont
  {{Scoccimarro}}, \citenamefont {{Zaldarriaga}},\ and\ \citenamefont
  {{Hui}}}]{Scoccimarro99}%
  \BibitemOpen
  \bibfield  {author} {\bibinfo {author} {\bibfnamefont {R.}~\bibnamefont
  {{Scoccimarro}}}, \bibinfo {author} {\bibfnamefont {M.}~\bibnamefont
  {{Zaldarriaga}}}, \ and\ \bibinfo {author} {\bibfnamefont {L.}~\bibnamefont
  {{Hui}}},\ }\href {\doibase 10.1086/308059} {\bibfield  {journal} {\bibinfo
  {journal} {\bibapj}\ }\textbf {\bibinfo {volume} {527}},\ \bibinfo {pages}
  {1} (\bibinfo {year} {1999})},\ \Eprint
  {http://arxiv.org/abs/astro-ph/9901099} {astro-ph/9901099} \BibitemShut
  {NoStop}%
\bibitem [{\citenamefont {{Rimes}}\ and\ \citenamefont
  {{Hamilton}}(2005)}]{Rimes05}%
  \BibitemOpen
  \bibfield  {author} {\bibinfo {author} {\bibfnamefont {C.~D.}\ \bibnamefont
  {{Rimes}}}\ and\ \bibinfo {author} {\bibfnamefont {A.~J.~S.}\ \bibnamefont
  {{Hamilton}}},\ }\href {\doibase 10.1111/j.1745-3933.2005.00051.x} {\bibfield
   {journal} {\bibinfo  {journal} {\bibmnras}\ }\textbf {\bibinfo {volume}
  {360}},\ \bibinfo {pages} {L82} (\bibinfo {year} {2005})},\ \Eprint
  {http://arxiv.org/abs/astro-ph/0502081} {astro-ph/0502081} \BibitemShut
  {NoStop}%
\bibitem [{\citenamefont {{Hamilton}}\ \emph {et~al.}(2006)\citenamefont
  {{Hamilton}}, \citenamefont {{Rimes}},\ and\ \citenamefont
  {{Scoccimarro}}}]{Hamilton06}%
  \BibitemOpen
  \bibfield  {author} {\bibinfo {author} {\bibfnamefont {A.~J.~S.}\
  \bibnamefont {{Hamilton}}}, \bibinfo {author} {\bibfnamefont {C.~D.}\
  \bibnamefont {{Rimes}}}, \ and\ \bibinfo {author} {\bibfnamefont
  {R.}~\bibnamefont {{Scoccimarro}}},\ }\href {\doibase
  10.1111/j.1365-2966.2006.10709.x} {\bibfield  {journal} {\bibinfo  {journal}
  {\bibmnras}\ }\textbf {\bibinfo {volume} {371}},\ \bibinfo {pages} {1188}
  (\bibinfo {year} {2006})},\ \Eprint {http://arxiv.org/abs/astro-ph/0511416}
  {astro-ph/0511416} \BibitemShut {NoStop}%
\bibitem [{\citenamefont {{Rimes}}\ and\ \citenamefont
  {{Hamilton}}(2006)}]{Rimes06}%
  \BibitemOpen
  \bibfield  {author} {\bibinfo {author} {\bibfnamefont {C.~D.}\ \bibnamefont
  {{Rimes}}}\ and\ \bibinfo {author} {\bibfnamefont {A.~J.~S.}\ \bibnamefont
  {{Hamilton}}},\ }\href {\doibase 10.1111/j.1365-2966.2006.10710.x} {\bibfield
   {journal} {\bibinfo  {journal} {\bibmnras}\ }\textbf {\bibinfo {volume}
  {371}},\ \bibinfo {pages} {1205} (\bibinfo {year} {2006})},\ \Eprint
  {http://arxiv.org/abs/astro-ph/0511418} {astro-ph/0511418} \BibitemShut
  {NoStop}%
\bibitem [{\citenamefont {{Neyrinck}}\ \emph {et~al.}(2006)\citenamefont
  {{Neyrinck}}, \citenamefont {{Szapudi}},\ and\ \citenamefont
  {{Rimes}}}]{Neyrinck06}%
  \BibitemOpen
  \bibfield  {author} {\bibinfo {author} {\bibfnamefont {M.~C.}\ \bibnamefont
  {{Neyrinck}}}, \bibinfo {author} {\bibfnamefont {I.}~\bibnamefont
  {{Szapudi}}}, \ and\ \bibinfo {author} {\bibfnamefont {C.~D.}\ \bibnamefont
  {{Rimes}}},\ }\href {\doibase 10.1111/j.1745-3933.2006.00190.x} {\bibfield
  {journal} {\bibinfo  {journal} {\bibmnras}\ }\textbf {\bibinfo {volume}
  {370}},\ \bibinfo {pages} {L66} (\bibinfo {year} {2006})},\ \Eprint
  {http://arxiv.org/abs/astro-ph/0604282} {astro-ph/0604282} \BibitemShut
  {NoStop}%
\bibitem [{\citenamefont {{Neyrinck}}\ and\ \citenamefont
  {{Szapudi}}(2007)}]{Neyrinck07}%
  \BibitemOpen
  \bibfield  {author} {\bibinfo {author} {\bibfnamefont {M.~C.}\ \bibnamefont
  {{Neyrinck}}}\ and\ \bibinfo {author} {\bibfnamefont {I.}~\bibnamefont
  {{Szapudi}}},\ }\href {\doibase 10.1111/j.1745-3933.2006.00275.x} {\bibfield
  {journal} {\bibinfo  {journal} {\bibmnras}\ }\textbf {\bibinfo {volume}
  {375}},\ \bibinfo {pages} {L51} (\bibinfo {year} {2007})},\ \Eprint
  {http://arxiv.org/abs/astro-ph/0610211} {astro-ph/0610211} \BibitemShut
  {NoStop}%
\bibitem [{\citenamefont {{Lee}}\ and\ \citenamefont {{Pen}}(2008)}]{Lee08}%
  \BibitemOpen
  \bibfield  {author} {\bibinfo {author} {\bibfnamefont {J.}~\bibnamefont
  {{Lee}}}\ and\ \bibinfo {author} {\bibfnamefont {U.-L.}\ \bibnamefont
  {{Pen}}},\ }\href {\doibase 10.1086/592820} {\bibfield  {journal} {\bibinfo
  {journal} {\bibapjl}\ }\textbf {\bibinfo {volume} {686}},\ \bibinfo {eid}
  {L1} (\bibinfo {year} {2008})},\ \Eprint {http://arxiv.org/abs/0807.1538}
  {arXiv:0807.1538} \BibitemShut {NoStop}%
\bibitem [{\citenamefont {{Semboloni}}\ \emph {et~al.}(2007)\citenamefont
  {{Semboloni}}, \citenamefont {{van Waerbeke}}, \citenamefont {{Heymans}},
  \citenamefont {{Hamana}}, \citenamefont {{Colombi}}, \citenamefont
  {{White}},\ and\ \citenamefont {{Mellier}}}]{Semboloni07}%
  \BibitemOpen
  \bibfield  {author} {\bibinfo {author} {\bibfnamefont {E.}~\bibnamefont
  {{Semboloni}}}, \bibinfo {author} {\bibfnamefont {L.}~\bibnamefont {{van
  Waerbeke}}}, \bibinfo {author} {\bibfnamefont {C.}~\bibnamefont {{Heymans}}},
  \bibinfo {author} {\bibfnamefont {T.}~\bibnamefont {{Hamana}}}, \bibinfo
  {author} {\bibfnamefont {S.}~\bibnamefont {{Colombi}}}, \bibinfo {author}
  {\bibfnamefont {M.}~\bibnamefont {{White}}}, \ and\ \bibinfo {author}
  {\bibfnamefont {Y.}~\bibnamefont {{Mellier}}},\ }\href {\doibase
  10.1111/j.1745-3933.2006.00266.x} {\bibfield  {journal} {\bibinfo  {journal}
  {\bibmnras}\ }\textbf {\bibinfo {volume} {375}},\ \bibinfo {pages} {L6}
  (\bibinfo {year} {2007})},\ \Eprint {http://arxiv.org/abs/astro-ph/0606648}
  {astro-ph/0606648} \BibitemShut {NoStop}%
\bibitem [{\citenamefont {{Takada}}\ and\ \citenamefont
  {{Jain}}(2009)}]{Takada09}%
  \BibitemOpen
  \bibfield  {author} {\bibinfo {author} {\bibfnamefont {M.}~\bibnamefont
  {{Takada}}}\ and\ \bibinfo {author} {\bibfnamefont {B.}~\bibnamefont
  {{Jain}}},\ }\href {\doibase 10.1111/j.1365-2966.2009.14504.x} {\bibfield
  {journal} {\bibinfo  {journal} {\bibmnras}\ }\textbf {\bibinfo {volume}
  {395}},\ \bibinfo {pages} {2065} (\bibinfo {year} {2009})},\ \Eprint
  {http://arxiv.org/abs/0810.4170} {arXiv:0810.4170} \BibitemShut {NoStop}%
\bibitem [{\citenamefont {{Pielorz}}\ \emph {et~al.}(2010)\citenamefont
  {{Pielorz}}, \citenamefont {{R{\"o}diger}}, \citenamefont {{Tereno}},\ and\
  \citenamefont {{Schneider}}}]{Pielorz10}%
  \BibitemOpen
  \bibfield  {author} {\bibinfo {author} {\bibfnamefont {J.}~\bibnamefont
  {{Pielorz}}}, \bibinfo {author} {\bibfnamefont {J.}~\bibnamefont
  {{R{\"o}diger}}}, \bibinfo {author} {\bibfnamefont {I.}~\bibnamefont
  {{Tereno}}}, \ and\ \bibinfo {author} {\bibfnamefont {P.}~\bibnamefont
  {{Schneider}}},\ }\href {\doibase 10.1051/0004-6361/200912854} {\bibfield
  {journal} {\bibinfo  {journal} {\bibaap}\ }\textbf {\bibinfo {volume}
  {514}},\ \bibinfo {eid} {A79} (\bibinfo {year} {2010})},\ \Eprint
  {http://arxiv.org/abs/0907.1524} {arXiv:0907.1524} \BibitemShut {NoStop}%
\bibitem [{\citenamefont {{Kiessling}}\ \emph {et~al.}(2011)\citenamefont
  {{Kiessling}}, \citenamefont {{Taylor}},\ and\ \citenamefont
  {{Heavens}}}]{Kiessling11b}%
  \BibitemOpen
  \bibfield  {author} {\bibinfo {author} {\bibfnamefont {A.}~\bibnamefont
  {{Kiessling}}}, \bibinfo {author} {\bibfnamefont {A.~N.}\ \bibnamefont
  {{Taylor}}}, \ and\ \bibinfo {author} {\bibfnamefont {A.~F.}\ \bibnamefont
  {{Heavens}}},\ }\href {\doibase 10.1111/j.1365-2966.2011.19108.x} {\bibfield
  {journal} {\bibinfo  {journal} {\bibmnras}\ }\textbf {\bibinfo {volume}
  {416}},\ \bibinfo {pages} {1045} (\bibinfo {year} {2011})},\ \Eprint
  {http://arxiv.org/abs/1103.3245} {arXiv:1103.3245} \BibitemShut {NoStop}%
\bibitem [{\citenamefont {{Harnois-D{\'e}raps}}\ and\ \citenamefont {{van
  Waerbeke}}(2015)}]{Harnois-Deraps15}%
  \BibitemOpen
  \bibfield  {author} {\bibinfo {author} {\bibfnamefont {J.}~\bibnamefont
  {{Harnois-D{\'e}raps}}}\ and\ \bibinfo {author} {\bibfnamefont
  {L.}~\bibnamefont {{van Waerbeke}}},\ }\href {\doibase 10.1093/mnras/stv794}
  {\bibfield  {journal} {\bibinfo  {journal} {\bibmnras}\ }\textbf {\bibinfo
  {volume} {450}},\ \bibinfo {pages} {2857} (\bibinfo {year} {2015})},\ \Eprint
  {http://arxiv.org/abs/1406.0543} {arXiv:1406.0543} \BibitemShut {NoStop}%
\bibitem [{\citenamefont {{Dodelson}}\ and\ \citenamefont
  {{Schneider}}(2013)}]{Dodelson13}%
  \BibitemOpen
  \bibfield  {author} {\bibinfo {author} {\bibfnamefont {S.}~\bibnamefont
  {{Dodelson}}}\ and\ \bibinfo {author} {\bibfnamefont {M.~D.}\ \bibnamefont
  {{Schneider}}},\ }\href {\doibase 10.1103/PhysRevD.88.063537} {\bibfield
  {journal} {\bibinfo  {journal} {\bibprd}\ }\textbf {\bibinfo {volume} {88}},\
  \bibinfo {eid} {063537} (\bibinfo {year} {2013})},\ \Eprint
  {http://arxiv.org/abs/1304.2593} {arXiv:1304.2593 [astro-ph.CO]} \BibitemShut
  {NoStop}%
\bibitem [{\citenamefont {{Taylor}}\ \emph {et~al.}(2013)\citenamefont
  {{Taylor}}, \citenamefont {{Joachimi}},\ and\ \citenamefont
  {{Kitching}}}]{Taylor13}%
  \BibitemOpen
  \bibfield  {author} {\bibinfo {author} {\bibfnamefont {A.}~\bibnamefont
  {{Taylor}}}, \bibinfo {author} {\bibfnamefont {B.}~\bibnamefont
  {{Joachimi}}}, \ and\ \bibinfo {author} {\bibfnamefont {T.}~\bibnamefont
  {{Kitching}}},\ }\href {\doibase 10.1093/mnras/stt270} {\bibfield  {journal}
  {\bibinfo  {journal} {\bibmnras}\ }\textbf {\bibinfo {volume} {432}},\
  \bibinfo {pages} {1928} (\bibinfo {year} {2013})},\ \Eprint
  {http://arxiv.org/abs/1212.4359} {arXiv:1212.4359} \BibitemShut {NoStop}%
\bibitem [{\citenamefont {{Percival}}\ \emph {et~al.}(2014)\citenamefont
  {{Percival}}, \citenamefont {{Ross}}, \citenamefont {{S{\'a}nchez}},
  \citenamefont {{Samushia}}, \citenamefont {{Burden}}, \citenamefont
  {{Crittenden}}, \citenamefont {{Cuesta}}, \citenamefont {{Magana}},
  \citenamefont {{Manera}}, \citenamefont {{Beutler}}, \citenamefont
  {{Chuang}}, \citenamefont {{Eisenstein}}, \citenamefont {{Ho}}, \citenamefont
  {{McBride}}, \citenamefont {{Montesano}}, \citenamefont {{Padmanabhan}},
  \citenamefont {{Reid}}, \citenamefont {{Saito}}, \citenamefont {{Schneider}},
  \citenamefont {{Seo}}, \citenamefont {{Tojeiro}},\ and\ \citenamefont
  {{Weaver}}}]{Percival14}%
  \BibitemOpen
  \bibfield  {author} {\bibinfo {author} {\bibfnamefont {W.~J.}\ \bibnamefont
  {{Percival}}}, \bibinfo {author} {\bibfnamefont {A.~J.}\ \bibnamefont
  {{Ross}}}, \bibinfo {author} {\bibfnamefont {A.~G.}\ \bibnamefont
  {{S{\'a}nchez}}}, \bibinfo {author} {\bibfnamefont {L.}~\bibnamefont
  {{Samushia}}}, \bibinfo {author} {\bibfnamefont {A.}~\bibnamefont
  {{Burden}}}, \bibinfo {author} {\bibfnamefont {R.}~\bibnamefont
  {{Crittenden}}}, \bibinfo {author} {\bibfnamefont {A.~J.}\ \bibnamefont
  {{Cuesta}}}, \bibinfo {author} {\bibfnamefont {M.~V.}\ \bibnamefont
  {{Magana}}}, \bibinfo {author} {\bibfnamefont {M.}~\bibnamefont {{Manera}}},
  \bibinfo {author} {\bibfnamefont {F.}~\bibnamefont {{Beutler}}}, \bibinfo
  {author} {\bibfnamefont {C.-H.}\ \bibnamefont {{Chuang}}}, \bibinfo {author}
  {\bibfnamefont {D.~J.}\ \bibnamefont {{Eisenstein}}}, \bibinfo {author}
  {\bibfnamefont {S.}~\bibnamefont {{Ho}}}, \bibinfo {author} {\bibfnamefont
  {C.~K.}\ \bibnamefont {{McBride}}}, \bibinfo {author} {\bibfnamefont
  {F.}~\bibnamefont {{Montesano}}}, \bibinfo {author} {\bibfnamefont
  {N.}~\bibnamefont {{Padmanabhan}}}, \bibinfo {author} {\bibfnamefont
  {B.}~\bibnamefont {{Reid}}}, \bibinfo {author} {\bibfnamefont
  {S.}~\bibnamefont {{Saito}}}, \bibinfo {author} {\bibfnamefont {D.~P.}\
  \bibnamefont {{Schneider}}}, \bibinfo {author} {\bibfnamefont {H.-J.}\
  \bibnamefont {{Seo}}}, \bibinfo {author} {\bibfnamefont {R.}~\bibnamefont
  {{Tojeiro}}}, \ and\ \bibinfo {author} {\bibfnamefont {B.~A.}\ \bibnamefont
  {{Weaver}}},\ }\href {\doibase 10.1093/mnras/stu112} {\bibfield  {journal}
  {\bibinfo  {journal} {\bibmnras}\ }\textbf {\bibinfo {volume} {439}},\
  \bibinfo {pages} {2531} (\bibinfo {year} {2014})},\ \Eprint
  {http://arxiv.org/abs/1312.4841} {arXiv:1312.4841} \BibitemShut {NoStop}%
\bibitem [{\citenamefont {{Harnois-D{\'e}raps}}\ \emph
  {et~al.}(2012)\citenamefont {{Harnois-D{\'e}raps}}, \citenamefont
  {{Vafaei}},\ and\ \citenamefont {{Van Waerbeke}}}]{Harnois-Deraps12}%
  \BibitemOpen
  \bibfield  {author} {\bibinfo {author} {\bibfnamefont {J.}~\bibnamefont
  {{Harnois-D{\'e}raps}}}, \bibinfo {author} {\bibfnamefont {S.}~\bibnamefont
  {{Vafaei}}}, \ and\ \bibinfo {author} {\bibfnamefont {L.}~\bibnamefont {{Van
  Waerbeke}}},\ }\href {\doibase 10.1111/j.1365-2966.2012.21624.x} {\bibfield
  {journal} {\bibinfo  {journal} {\bibmnras}\ }\textbf {\bibinfo {volume}
  {426}},\ \bibinfo {pages} {1262} (\bibinfo {year} {2012})},\ \Eprint
  {http://arxiv.org/abs/1202.2332} {arXiv:1202.2332} \BibitemShut {NoStop}%
\bibitem [{\citenamefont {{Blot}}\ \emph
  {et~al.}(2015{\natexlab{a}})\citenamefont {{Blot}}, \citenamefont
  {{Corasaniti}}, \citenamefont {{Alimi}}, \citenamefont {{Reverdy}},\ and\
  \citenamefont {{Rasera}}}]{Blot15}%
  \BibitemOpen
  \bibfield  {author} {\bibinfo {author} {\bibfnamefont {L.}~\bibnamefont
  {{Blot}}}, \bibinfo {author} {\bibfnamefont {P.~S.}\ \bibnamefont
  {{Corasaniti}}}, \bibinfo {author} {\bibfnamefont {J.-M.}\ \bibnamefont
  {{Alimi}}}, \bibinfo {author} {\bibfnamefont {V.}~\bibnamefont {{Reverdy}}},
  \ and\ \bibinfo {author} {\bibfnamefont {Y.}~\bibnamefont {{Rasera}}},\
  }\href {\doibase 10.1093/mnras/stu2190} {\bibfield  {journal} {\bibinfo
  {journal} {\bibmnras}\ }\textbf {\bibinfo {volume} {446}},\ \bibinfo {pages}
  {1756} (\bibinfo {year} {2015}{\natexlab{a}})},\ \Eprint
  {http://arxiv.org/abs/1406.2713} {arXiv:1406.2713} \BibitemShut {NoStop}%
\bibitem [{\citenamefont {{Blot}}\ \emph
  {et~al.}(2015{\natexlab{b}})\citenamefont {{Blot}}, \citenamefont
  {{Corasaniti}}, \citenamefont {{Amendola}},\ and\ \citenamefont
  {{Kitching}}}]{Blot15arxiv}%
  \BibitemOpen
  \bibfield  {author} {\bibinfo {author} {\bibfnamefont {L.}~\bibnamefont
  {{Blot}}}, \bibinfo {author} {\bibfnamefont {P.~S.}\ \bibnamefont
  {{Corasaniti}}}, \bibinfo {author} {\bibfnamefont {L.}~\bibnamefont
  {{Amendola}}}, \ and\ \bibinfo {author} {\bibfnamefont {T.~D.}\ \bibnamefont
  {{Kitching}}},\ }\href@noop {} {\bibfield  {journal} {\bibinfo  {journal}
  {ArXiv e-prints}\ } (\bibinfo {year} {2015}{\natexlab{b}})},\ \Eprint
  {http://arxiv.org/abs/1512.05383} {arXiv:1512.05383} \BibitemShut {NoStop}%
\bibitem [{\citenamefont {{Coles}}\ and\ \citenamefont
  {{Jones}}(1991)}]{Coles91}%
  \BibitemOpen
  \bibfield  {author} {\bibinfo {author} {\bibfnamefont {P.}~\bibnamefont
  {{Coles}}}\ and\ \bibinfo {author} {\bibfnamefont {B.}~\bibnamefont
  {{Jones}}},\ }\href@noop {} {\bibfield  {journal} {\bibinfo  {journal}
  {\bibmnras}\ }\textbf {\bibinfo {volume} {248}},\ \bibinfo {pages} {1}
  (\bibinfo {year} {1991})}\BibitemShut {NoStop}%
\bibitem [{\citenamefont {{Cole}}\ \emph {et~al.}(2005)\citenamefont {{Cole}},
  \citenamefont {{Percival}}, \citenamefont {{Peacock}}, \citenamefont
  {{Norberg}}, \citenamefont {{Baugh}}, \citenamefont {{Frenk}}, \citenamefont
  {{Baldry}}, \citenamefont {{Bland-Hawthorn}}, \citenamefont {{Bridges}},
  \citenamefont {{Cannon}}, \citenamefont {{Colless}}, \citenamefont
  {{Collins}}, \citenamefont {{Couch}}, \citenamefont {{Cross}}, \citenamefont
  {{Dalton}}, \citenamefont {{Eke}}, \citenamefont {{De Propris}},
  \citenamefont {{Driver}}, \citenamefont {{Efstathiou}}, \citenamefont
  {{Ellis}}, \citenamefont {{Glazebrook}}, \citenamefont {{Jackson}},
  \citenamefont {{Jenkins}}, \citenamefont {{Lahav}}, \citenamefont {{Lewis}},
  \citenamefont {{Lumsden}}, \citenamefont {{Maddox}}, \citenamefont
  {{Madgwick}}, \citenamefont {{Peterson}}, \citenamefont {{Sutherland}},\ and\
  \citenamefont {{Taylor}}}]{Cole05}%
  \BibitemOpen
  \bibfield  {author} {\bibinfo {author} {\bibfnamefont {S.}~\bibnamefont
  {{Cole}}}, \bibinfo {author} {\bibfnamefont {W.~J.}\ \bibnamefont
  {{Percival}}}, \bibinfo {author} {\bibfnamefont {J.~A.}\ \bibnamefont
  {{Peacock}}}, \bibinfo {author} {\bibfnamefont {P.}~\bibnamefont
  {{Norberg}}}, \bibinfo {author} {\bibfnamefont {C.~M.}\ \bibnamefont
  {{Baugh}}}, \bibinfo {author} {\bibfnamefont {C.~S.}\ \bibnamefont
  {{Frenk}}}, \bibinfo {author} {\bibfnamefont {I.}~\bibnamefont {{Baldry}}},
  \bibinfo {author} {\bibfnamefont {J.}~\bibnamefont {{Bland-Hawthorn}}},
  \bibinfo {author} {\bibfnamefont {T.}~\bibnamefont {{Bridges}}}, \bibinfo
  {author} {\bibfnamefont {R.}~\bibnamefont {{Cannon}}}, \bibinfo {author}
  {\bibfnamefont {M.}~\bibnamefont {{Colless}}}, \bibinfo {author}
  {\bibfnamefont {C.}~\bibnamefont {{Collins}}}, \bibinfo {author}
  {\bibfnamefont {W.}~\bibnamefont {{Couch}}}, \bibinfo {author} {\bibfnamefont
  {N.~J.~G.}\ \bibnamefont {{Cross}}}, \bibinfo {author} {\bibfnamefont
  {G.}~\bibnamefont {{Dalton}}}, \bibinfo {author} {\bibfnamefont {V.~R.}\
  \bibnamefont {{Eke}}}, \bibinfo {author} {\bibfnamefont {R.}~\bibnamefont
  {{De Propris}}}, \bibinfo {author} {\bibfnamefont {S.~P.}\ \bibnamefont
  {{Driver}}}, \bibinfo {author} {\bibfnamefont {G.}~\bibnamefont
  {{Efstathiou}}}, \bibinfo {author} {\bibfnamefont {R.~S.}\ \bibnamefont
  {{Ellis}}}, \bibinfo {author} {\bibfnamefont {K.}~\bibnamefont
  {{Glazebrook}}}, \bibinfo {author} {\bibfnamefont {C.}~\bibnamefont
  {{Jackson}}}, \bibinfo {author} {\bibfnamefont {A.}~\bibnamefont
  {{Jenkins}}}, \bibinfo {author} {\bibfnamefont {O.}~\bibnamefont {{Lahav}}},
  \bibinfo {author} {\bibfnamefont {I.}~\bibnamefont {{Lewis}}}, \bibinfo
  {author} {\bibfnamefont {S.}~\bibnamefont {{Lumsden}}}, \bibinfo {author}
  {\bibfnamefont {S.}~\bibnamefont {{Maddox}}}, \bibinfo {author}
  {\bibfnamefont {D.}~\bibnamefont {{Madgwick}}}, \bibinfo {author}
  {\bibfnamefont {B.~A.}\ \bibnamefont {{Peterson}}}, \bibinfo {author}
  {\bibfnamefont {W.}~\bibnamefont {{Sutherland}}}, \ and\ \bibinfo {author}
  {\bibfnamefont {K.}~\bibnamefont {{Taylor}}},\ }\href {\doibase
  10.1111/j.1365-2966.2005.09318.x} {\bibfield  {journal} {\bibinfo  {journal}
  {\bibmnras}\ }\textbf {\bibinfo {volume} {362}},\ \bibinfo {pages} {505}
  (\bibinfo {year} {2005})},\ \Eprint {http://arxiv.org/abs/astro-ph/0501174}
  {astro-ph/0501174} \BibitemShut {NoStop}%
\bibitem [{\citenamefont {{Scoccimarro}}\ and\ \citenamefont
  {{Sheth}}(2002)}]{Scoccimarro02}%
  \BibitemOpen
  \bibfield  {author} {\bibinfo {author} {\bibfnamefont {R.}~\bibnamefont
  {{Scoccimarro}}}\ and\ \bibinfo {author} {\bibfnamefont {R.~K.}\ \bibnamefont
  {{Sheth}}},\ }\href {\doibase 10.1046/j.1365-8711.2002.04999.x} {\bibfield
  {journal} {\bibinfo  {journal} {\bibmnras}\ }\textbf {\bibinfo {volume}
  {329}},\ \bibinfo {pages} {629} (\bibinfo {year} {2002})},\ \Eprint
  {http://arxiv.org/abs/astro-ph/0106120} {astro-ph/0106120} \BibitemShut
  {NoStop}%
\bibitem [{\citenamefont {{Manera}}\ \emph {et~al.}(2013)\citenamefont
  {{Manera}}, \citenamefont {{Scoccimarro}}, \citenamefont {{Percival}},
  \citenamefont {{Samushia}}, \citenamefont {{McBride}}, \citenamefont
  {{Ross}}, \citenamefont {{Sheth}}, \citenamefont {{White}}, \citenamefont
  {{Reid}}, \citenamefont {{S{\'a}nchez}}, \citenamefont {{de Putter}},
  \citenamefont {{Xu}}, \citenamefont {{Berlind}}, \citenamefont {{Brinkmann}},
  \citenamefont {{Maraston}}, \citenamefont {{Nichol}}, \citenamefont
  {{Montesano}}, \citenamefont {{Padmanabhan}}, \citenamefont {{Skibba}},
  \citenamefont {{Tojeiro}},\ and\ \citenamefont {{Weaver}}}]{Manera13}%
  \BibitemOpen
  \bibfield  {author} {\bibinfo {author} {\bibfnamefont {M.}~\bibnamefont
  {{Manera}}}, \bibinfo {author} {\bibfnamefont {R.}~\bibnamefont
  {{Scoccimarro}}}, \bibinfo {author} {\bibfnamefont {W.~J.}\ \bibnamefont
  {{Percival}}}, \bibinfo {author} {\bibfnamefont {L.}~\bibnamefont
  {{Samushia}}}, \bibinfo {author} {\bibfnamefont {C.~K.}\ \bibnamefont
  {{McBride}}}, \bibinfo {author} {\bibfnamefont {A.~J.}\ \bibnamefont
  {{Ross}}}, \bibinfo {author} {\bibfnamefont {R.~K.}\ \bibnamefont {{Sheth}}},
  \bibinfo {author} {\bibfnamefont {M.}~\bibnamefont {{White}}}, \bibinfo
  {author} {\bibfnamefont {B.~A.}\ \bibnamefont {{Reid}}}, \bibinfo {author}
  {\bibfnamefont {A.~G.}\ \bibnamefont {{S{\'a}nchez}}}, \bibinfo {author}
  {\bibfnamefont {R.}~\bibnamefont {{de Putter}}}, \bibinfo {author}
  {\bibfnamefont {X.}~\bibnamefont {{Xu}}}, \bibinfo {author} {\bibfnamefont
  {A.~A.}\ \bibnamefont {{Berlind}}}, \bibinfo {author} {\bibfnamefont
  {J.}~\bibnamefont {{Brinkmann}}}, \bibinfo {author} {\bibfnamefont
  {C.}~\bibnamefont {{Maraston}}}, \bibinfo {author} {\bibfnamefont
  {B.}~\bibnamefont {{Nichol}}}, \bibinfo {author} {\bibfnamefont
  {F.}~\bibnamefont {{Montesano}}}, \bibinfo {author} {\bibfnamefont
  {N.}~\bibnamefont {{Padmanabhan}}}, \bibinfo {author} {\bibfnamefont {R.~A.}\
  \bibnamefont {{Skibba}}}, \bibinfo {author} {\bibfnamefont {R.}~\bibnamefont
  {{Tojeiro}}}, \ and\ \bibinfo {author} {\bibfnamefont {B.~A.}\ \bibnamefont
  {{Weaver}}},\ }\href {\doibase 10.1093/mnras/sts084} {\bibfield  {journal}
  {\bibinfo  {journal} {\bibmnras}\ }\textbf {\bibinfo {volume} {428}},\
  \bibinfo {pages} {1036} (\bibinfo {year} {2013})},\ \Eprint
  {http://arxiv.org/abs/1203.6609} {arXiv:1203.6609} \BibitemShut {NoStop}%
\bibitem [{\citenamefont {{Manera}}\ \emph {et~al.}(2015)\citenamefont
  {{Manera}}, \citenamefont {{Samushia}}, \citenamefont {{Tojeiro}},
  \citenamefont {{Howlett}}, \citenamefont {{Ross}}, \citenamefont
  {{Percival}}, \citenamefont {{Gil-Mar{\'{\i}}n}}, \citenamefont
  {{Brownstein}}, \citenamefont {{Burden}},\ and\ \citenamefont
  {{Montesano}}}]{Manera15}%
  \BibitemOpen
  \bibfield  {author} {\bibinfo {author} {\bibfnamefont {M.}~\bibnamefont
  {{Manera}}}, \bibinfo {author} {\bibfnamefont {L.}~\bibnamefont
  {{Samushia}}}, \bibinfo {author} {\bibfnamefont {R.}~\bibnamefont
  {{Tojeiro}}}, \bibinfo {author} {\bibfnamefont {C.}~\bibnamefont
  {{Howlett}}}, \bibinfo {author} {\bibfnamefont {A.~J.}\ \bibnamefont
  {{Ross}}}, \bibinfo {author} {\bibfnamefont {W.~J.}\ \bibnamefont
  {{Percival}}}, \bibinfo {author} {\bibfnamefont {H.}~\bibnamefont
  {{Gil-Mar{\'{\i}}n}}}, \bibinfo {author} {\bibfnamefont {J.~R.}\ \bibnamefont
  {{Brownstein}}}, \bibinfo {author} {\bibfnamefont {A.}~\bibnamefont
  {{Burden}}}, \ and\ \bibinfo {author} {\bibfnamefont {F.}~\bibnamefont
  {{Montesano}}},\ }\href {\doibase 10.1093/mnras/stu2465} {\bibfield
  {journal} {\bibinfo  {journal} {\bibmnras}\ }\textbf {\bibinfo {volume}
  {447}},\ \bibinfo {pages} {437} (\bibinfo {year} {2015})},\ \Eprint
  {http://arxiv.org/abs/1401.4171} {arXiv:1401.4171} \BibitemShut {NoStop}%
\bibitem [{\citenamefont {{Monaco}}\ \emph {et~al.}(2002)\citenamefont
  {{Monaco}}, \citenamefont {{Theuns}},\ and\ \citenamefont
  {{Taffoni}}}]{Monaco02}%
  \BibitemOpen
  \bibfield  {author} {\bibinfo {author} {\bibfnamefont {P.}~\bibnamefont
  {{Monaco}}}, \bibinfo {author} {\bibfnamefont {T.}~\bibnamefont {{Theuns}}},
  \ and\ \bibinfo {author} {\bibfnamefont {G.}~\bibnamefont {{Taffoni}}},\
  }\href {\doibase 10.1046/j.1365-8711.2002.05162.x} {\bibfield  {journal}
  {\bibinfo  {journal} {\bibmnras}\ }\textbf {\bibinfo {volume} {331}},\
  \bibinfo {pages} {587} (\bibinfo {year} {2002})},\ \Eprint
  {http://arxiv.org/abs/astro-ph/0109323} {astro-ph/0109323} \BibitemShut
  {NoStop}%
\bibitem [{\citenamefont {{Monaco}}\ \emph {et~al.}(2013)\citenamefont
  {{Monaco}}, \citenamefont {{Sefusatti}}, \citenamefont {{Borgani}},
  \citenamefont {{Crocce}}, \citenamefont {{Fosalba}}, \citenamefont
  {{Sheth}},\ and\ \citenamefont {{Theuns}}}]{Monaco13}%
  \BibitemOpen
  \bibfield  {author} {\bibinfo {author} {\bibfnamefont {P.}~\bibnamefont
  {{Monaco}}}, \bibinfo {author} {\bibfnamefont {E.}~\bibnamefont
  {{Sefusatti}}}, \bibinfo {author} {\bibfnamefont {S.}~\bibnamefont
  {{Borgani}}}, \bibinfo {author} {\bibfnamefont {M.}~\bibnamefont {{Crocce}}},
  \bibinfo {author} {\bibfnamefont {P.}~\bibnamefont {{Fosalba}}}, \bibinfo
  {author} {\bibfnamefont {R.~K.}\ \bibnamefont {{Sheth}}}, \ and\ \bibinfo
  {author} {\bibfnamefont {T.}~\bibnamefont {{Theuns}}},\ }\href {\doibase
  10.1093/mnras/stt907} {\bibfield  {journal} {\bibinfo  {journal} {\bibmnras}\
  }\textbf {\bibinfo {volume} {433}},\ \bibinfo {pages} {2389} (\bibinfo {year}
  {2013})},\ \Eprint {http://arxiv.org/abs/1305.1505} {arXiv:1305.1505}
  \BibitemShut {NoStop}%
\bibitem [{\citenamefont {{White}}\ \emph {et~al.}(2014)\citenamefont
  {{White}}, \citenamefont {{Tinker}},\ and\ \citenamefont
  {{McBride}}}]{White14}%
  \BibitemOpen
  \bibfield  {author} {\bibinfo {author} {\bibfnamefont {M.}~\bibnamefont
  {{White}}}, \bibinfo {author} {\bibfnamefont {J.~L.}\ \bibnamefont
  {{Tinker}}}, \ and\ \bibinfo {author} {\bibfnamefont {C.~K.}\ \bibnamefont
  {{McBride}}},\ }\href {\doibase 10.1093/mnras/stt2071} {\bibfield  {journal}
  {\bibinfo  {journal} {\bibmnras}\ }\textbf {\bibinfo {volume} {437}},\
  \bibinfo {pages} {2594} (\bibinfo {year} {2014})},\ \Eprint
  {http://arxiv.org/abs/1309.5532} {arXiv:1309.5532} \BibitemShut {NoStop}%
\bibitem [{\citenamefont {{Kitaura}}\ and\ \citenamefont
  {{He{\ss}}}(2013)}]{Kitaura13b}%
  \BibitemOpen
  \bibfield  {author} {\bibinfo {author} {\bibfnamefont {F.-S.}\ \bibnamefont
  {{Kitaura}}}\ and\ \bibinfo {author} {\bibfnamefont {S.}~\bibnamefont
  {{He{\ss}}}},\ }\href {\doibase 10.1093/mnrasl/slt101} {\bibfield  {journal}
  {\bibinfo  {journal} {\bibmnras}\ }\textbf {\bibinfo {volume} {435}},\
  \bibinfo {pages} {L78} (\bibinfo {year} {2013})},\ \Eprint
  {http://arxiv.org/abs/1212.3514} {arXiv:1212.3514} \BibitemShut {NoStop}%
\bibitem [{\citenamefont {{Chuang}}\ \emph {et~al.}(2015)\citenamefont
  {{Chuang}}, \citenamefont {{Kitaura}}, \citenamefont {{Prada}}, \citenamefont
  {{Zhao}},\ and\ \citenamefont {{Yepes}}}]{Chuang15}%
  \BibitemOpen
  \bibfield  {author} {\bibinfo {author} {\bibfnamefont {C.-H.}\ \bibnamefont
  {{Chuang}}}, \bibinfo {author} {\bibfnamefont {F.-S.}\ \bibnamefont
  {{Kitaura}}}, \bibinfo {author} {\bibfnamefont {F.}~\bibnamefont {{Prada}}},
  \bibinfo {author} {\bibfnamefont {C.}~\bibnamefont {{Zhao}}}, \ and\ \bibinfo
  {author} {\bibfnamefont {G.}~\bibnamefont {{Yepes}}},\ }\href {\doibase
  10.1093/mnras/stu2301} {\bibfield  {journal} {\bibinfo  {journal}
  {\bibmnras}\ }\textbf {\bibinfo {volume} {446}},\ \bibinfo {pages} {2621}
  (\bibinfo {year} {2015})},\ \Eprint {http://arxiv.org/abs/1409.1124}
  {arXiv:1409.1124} \BibitemShut {NoStop}%
\bibitem [{\citenamefont {{Tassev}}\ \emph {et~al.}(2013)\citenamefont
  {{Tassev}}, \citenamefont {{Zaldarriaga}},\ and\ \citenamefont
  {{Eisenstein}}}]{Tassev13}%
  \BibitemOpen
  \bibfield  {author} {\bibinfo {author} {\bibfnamefont {S.}~\bibnamefont
  {{Tassev}}}, \bibinfo {author} {\bibfnamefont {M.}~\bibnamefont
  {{Zaldarriaga}}}, \ and\ \bibinfo {author} {\bibfnamefont {D.~J.}\
  \bibnamefont {{Eisenstein}}},\ }\href {\doibase
  10.1088/1475-7516/2013/06/036} {\bibfield  {journal} {\bibinfo  {journal}
  {\bibjcap}\ }\textbf {\bibinfo {volume} {6}},\ \bibinfo {eid} {036} (\bibinfo
  {year} {2013})},\ \Eprint {http://arxiv.org/abs/1301.0322} {arXiv:1301.0322
  [astro-ph.CO]} \BibitemShut {NoStop}%
\bibitem [{\citenamefont {{Tassev}}\ \emph {et~al.}(2015)\citenamefont
  {{Tassev}}, \citenamefont {{Eisenstein}}, \citenamefont {{Wandelt}},\ and\
  \citenamefont {{Zaldarriaga}}}]{Tassev15}%
  \BibitemOpen
  \bibfield  {author} {\bibinfo {author} {\bibfnamefont {S.}~\bibnamefont
  {{Tassev}}}, \bibinfo {author} {\bibfnamefont {D.~J.}\ \bibnamefont
  {{Eisenstein}}}, \bibinfo {author} {\bibfnamefont {B.~D.}\ \bibnamefont
  {{Wandelt}}}, \ and\ \bibinfo {author} {\bibfnamefont {M.}~\bibnamefont
  {{Zaldarriaga}}},\ }\href@noop {} {\bibfield  {journal} {\bibinfo  {journal}
  {ArXiv e-prints}\ } (\bibinfo {year} {2015})},\ \Eprint
  {http://arxiv.org/abs/1502.07751} {arXiv:1502.07751} \BibitemShut {NoStop}%
\bibitem [{\citenamefont {{Koda}}\ \emph {et~al.}(2015)\citenamefont {{Koda}},
  \citenamefont {{Blake}}, \citenamefont {{Beutler}}, \citenamefont {{Kazin}},\
  and\ \citenamefont {{Marin}}}]{Koda15}%
  \BibitemOpen
  \bibfield  {author} {\bibinfo {author} {\bibfnamefont {J.}~\bibnamefont
  {{Koda}}}, \bibinfo {author} {\bibfnamefont {C.}~\bibnamefont {{Blake}}},
  \bibinfo {author} {\bibfnamefont {F.}~\bibnamefont {{Beutler}}}, \bibinfo
  {author} {\bibfnamefont {E.}~\bibnamefont {{Kazin}}}, \ and\ \bibinfo
  {author} {\bibfnamefont {F.}~\bibnamefont {{Marin}}},\ }\href@noop {}
  {\bibfield  {journal} {\bibinfo  {journal} {ArXiv e-prints}\ } (\bibinfo
  {year} {2015})},\ \Eprint {http://arxiv.org/abs/1507.05329}
  {arXiv:1507.05329} \BibitemShut {NoStop}%
\bibitem [{\citenamefont {{Howlett}}\ \emph {et~al.}(2015)\citenamefont
  {{Howlett}}, \citenamefont {{Manera}},\ and\ \citenamefont
  {{Percival}}}]{Howlett15}%
  \BibitemOpen
  \bibfield  {author} {\bibinfo {author} {\bibfnamefont {C.}~\bibnamefont
  {{Howlett}}}, \bibinfo {author} {\bibfnamefont {M.}~\bibnamefont {{Manera}}},
  \ and\ \bibinfo {author} {\bibfnamefont {W.~J.}\ \bibnamefont {{Percival}}},\
  }\href {\doibase 10.1016/j.ascom.2015.07.003} {\bibfield  {journal} {\bibinfo
   {journal} {Astronomy and Computing}\ }\textbf {\bibinfo {volume} {12}},\
  \bibinfo {pages} {109} (\bibinfo {year} {2015})},\ \Eprint
  {http://arxiv.org/abs/1506.03737} {arXiv:1506.03737} \BibitemShut {NoStop}%
\bibitem [{\citenamefont {{Izard}}\ \emph {et~al.}(2015)\citenamefont
  {{Izard}}, \citenamefont {{Crocce}},\ and\ \citenamefont
  {{Fosalba}}}]{Izard15arxiv}%
  \BibitemOpen
  \bibfield  {author} {\bibinfo {author} {\bibfnamefont {A.}~\bibnamefont
  {{Izard}}}, \bibinfo {author} {\bibfnamefont {M.}~\bibnamefont {{Crocce}}}, \
  and\ \bibinfo {author} {\bibfnamefont {P.}~\bibnamefont {{Fosalba}}},\
  }\href@noop {} {\bibfield  {journal} {\bibinfo  {journal} {ArXiv e-prints}\ }
  (\bibinfo {year} {2015})},\ \Eprint {http://arxiv.org/abs/1509.04685}
  {arXiv:1509.04685} \BibitemShut {NoStop}%
\bibitem [{\citenamefont {{Avila}}\ \emph {et~al.}(2015)\citenamefont
  {{Avila}}, \citenamefont {{Murray}}, \citenamefont {{Knebe}}, \citenamefont
  {{Power}}, \citenamefont {{Robotham}},\ and\ \citenamefont
  {{Garcia-Bellido}}}]{Avila15}%
  \BibitemOpen
  \bibfield  {author} {\bibinfo {author} {\bibfnamefont {S.}~\bibnamefont
  {{Avila}}}, \bibinfo {author} {\bibfnamefont {S.~G.}\ \bibnamefont
  {{Murray}}}, \bibinfo {author} {\bibfnamefont {A.}~\bibnamefont {{Knebe}}},
  \bibinfo {author} {\bibfnamefont {C.}~\bibnamefont {{Power}}}, \bibinfo
  {author} {\bibfnamefont {A.~S.~G.}\ \bibnamefont {{Robotham}}}, \ and\
  \bibinfo {author} {\bibfnamefont {J.}~\bibnamefont {{Garcia-Bellido}}},\
  }\href {\doibase 10.1093/mnras/stv711} {\bibfield  {journal} {\bibinfo
  {journal} {\bibmnras}\ }\textbf {\bibinfo {volume} {450}},\ \bibinfo {pages}
  {1856} (\bibinfo {year} {2015})},\ \Eprint {http://arxiv.org/abs/1412.5228}
  {arXiv:1412.5228} \BibitemShut {NoStop}%
\bibitem [{\citenamefont {{Feng}}\ \emph {et~al.}(2016)\citenamefont {{Feng}},
  \citenamefont {{Chu}},\ and\ \citenamefont {{Seljak}}}]{Feng16}%
  \BibitemOpen
  \bibfield  {author} {\bibinfo {author} {\bibfnamefont {Y.}~\bibnamefont
  {{Feng}}}, \bibinfo {author} {\bibfnamefont {M.-Y.}\ \bibnamefont {{Chu}}}, \
  and\ \bibinfo {author} {\bibfnamefont {U.}~\bibnamefont {{Seljak}}},\
  }\href@noop {} {\bibfield  {journal} {\bibinfo  {journal} {ArXiv e-prints}\ }
  (\bibinfo {year} {2016})},\ \Eprint {http://arxiv.org/abs/1603.00476}
  {arXiv:1603.00476} \BibitemShut {NoStop}%
\bibitem [{\citenamefont {{Seo}}\ \emph {et~al.}(2011)\citenamefont {{Seo}},
  \citenamefont {{Sato}}, \citenamefont {{Dodelson}}, \citenamefont {{Jain}},\
  and\ \citenamefont {{Takada}}}]{Seo11}%
  \BibitemOpen
  \bibfield  {author} {\bibinfo {author} {\bibfnamefont {H.-J.}\ \bibnamefont
  {{Seo}}}, \bibinfo {author} {\bibfnamefont {M.}~\bibnamefont {{Sato}}},
  \bibinfo {author} {\bibfnamefont {S.}~\bibnamefont {{Dodelson}}}, \bibinfo
  {author} {\bibfnamefont {B.}~\bibnamefont {{Jain}}}, \ and\ \bibinfo {author}
  {\bibfnamefont {M.}~\bibnamefont {{Takada}}},\ }\href {\doibase
  10.1088/2041-8205/729/1/L11} {\bibfield  {journal} {\bibinfo  {journal}
  {\bibapjl}\ }\textbf {\bibinfo {volume} {729}},\ \bibinfo {eid} {L11}
  (\bibinfo {year} {2011})},\ \Eprint {http://arxiv.org/abs/1008.0349}
  {arXiv:1008.0349 [astro-ph.CO]} \BibitemShut {NoStop}%
\bibitem [{\citenamefont {{Seo}}\ \emph {et~al.}(2012)\citenamefont {{Seo}},
  \citenamefont {{Sato}}, \citenamefont {{Takada}},\ and\ \citenamefont
  {{Dodelson}}}]{Seo12}%
  \BibitemOpen
  \bibfield  {author} {\bibinfo {author} {\bibfnamefont {H.-J.}\ \bibnamefont
  {{Seo}}}, \bibinfo {author} {\bibfnamefont {M.}~\bibnamefont {{Sato}}},
  \bibinfo {author} {\bibfnamefont {M.}~\bibnamefont {{Takada}}}, \ and\
  \bibinfo {author} {\bibfnamefont {S.}~\bibnamefont {{Dodelson}}},\ }\href
  {\doibase 10.1088/0004-637X/748/1/57} {\bibfield  {journal} {\bibinfo
  {journal} {\bibapj}\ }\textbf {\bibinfo {volume} {748}},\ \bibinfo {eid} {57}
  (\bibinfo {year} {2012})},\ \Eprint {http://arxiv.org/abs/1109.5639}
  {arXiv:1109.5639 [astro-ph.CO]} \BibitemShut {NoStop}%
\bibitem [{\citenamefont {{Carron}}\ \emph {et~al.}(2014)\citenamefont
  {{Carron}}, \citenamefont {{Wolk}},\ and\ \citenamefont
  {{Szapudi}}}]{Carron14b}%
  \BibitemOpen
  \bibfield  {author} {\bibinfo {author} {\bibfnamefont {J.}~\bibnamefont
  {{Carron}}}, \bibinfo {author} {\bibfnamefont {M.}~\bibnamefont {{Wolk}}}, \
  and\ \bibinfo {author} {\bibfnamefont {I.}~\bibnamefont {{Szapudi}}},\ }\href
  {\doibase 10.1093/mnras/stu1527} {\bibfield  {journal} {\bibinfo  {journal}
  {\bibmnras}\ }\textbf {\bibinfo {volume} {444}},\ \bibinfo {pages} {994}
  (\bibinfo {year} {2014})},\ \Eprint {http://arxiv.org/abs/1406.6072}
  {arXiv:1406.6072} \BibitemShut {NoStop}%
\bibitem [{\citenamefont {{Takahashi}}\ \emph {et~al.}(2014)\citenamefont
  {{Takahashi}}, \citenamefont {{Soma}}, \citenamefont {{Takada}},\ and\
  \citenamefont {{Kayo}}}]{Takahashi14}%
  \BibitemOpen
  \bibfield  {author} {\bibinfo {author} {\bibfnamefont {R.}~\bibnamefont
  {{Takahashi}}}, \bibinfo {author} {\bibfnamefont {S.}~\bibnamefont {{Soma}}},
  \bibinfo {author} {\bibfnamefont {M.}~\bibnamefont {{Takada}}}, \ and\
  \bibinfo {author} {\bibfnamefont {I.}~\bibnamefont {{Kayo}}},\ }\href
  {\doibase 10.1093/mnras/stu1693} {\bibfield  {journal} {\bibinfo  {journal}
  {\bibmnras}\ }\textbf {\bibinfo {volume} {444}},\ \bibinfo {pages} {3473}
  (\bibinfo {year} {2014})},\ \Eprint {http://arxiv.org/abs/1405.2666}
  {arXiv:1405.2666} \BibitemShut {NoStop}%
\bibitem [{\citenamefont {{Carron}}\ and\ \citenamefont
  {{Szapudi}}(2015)}]{Carron15}%
  \BibitemOpen
  \bibfield  {author} {\bibinfo {author} {\bibfnamefont {J.}~\bibnamefont
  {{Carron}}}\ and\ \bibinfo {author} {\bibfnamefont {I.}~\bibnamefont
  {{Szapudi}}},\ }\href {\doibase 10.1093/mnras/stu2501} {\bibfield  {journal}
  {\bibinfo  {journal} {\bibmnras}\ }\textbf {\bibinfo {volume} {447}},\
  \bibinfo {pages} {671} (\bibinfo {year} {2015})},\ \Eprint
  {http://arxiv.org/abs/1408.1744} {arXiv:1408.1744} \BibitemShut {NoStop}%
\bibitem [{\citenamefont {{Friedrich}}\ \emph {et~al.}(2016)\citenamefont
  {{Friedrich}}, \citenamefont {{Seitz}}, \citenamefont {{Eifler}},\ and\
  \citenamefont {{Gruen}}}]{Friedrich16}%
  \BibitemOpen
  \bibfield  {author} {\bibinfo {author} {\bibfnamefont {O.}~\bibnamefont
  {{Friedrich}}}, \bibinfo {author} {\bibfnamefont {S.}~\bibnamefont
  {{Seitz}}}, \bibinfo {author} {\bibfnamefont {T.~F.}\ \bibnamefont
  {{Eifler}}}, \ and\ \bibinfo {author} {\bibfnamefont {D.}~\bibnamefont
  {{Gruen}}},\ }\href {\doibase 10.1093/mnras/stv2833} {\bibfield  {journal}
  {\bibinfo  {journal} {\bibmnras}\ }\textbf {\bibinfo {volume} {456}},\
  \bibinfo {pages} {2662} (\bibinfo {year} {2016})},\ \Eprint
  {http://arxiv.org/abs/1508.00895} {arXiv:1508.00895} \BibitemShut {NoStop}%
\bibitem [{\citenamefont {{Colombi}}(1994)}]{Colombi94}%
  \BibitemOpen
  \bibfield  {author} {\bibinfo {author} {\bibfnamefont {S.}~\bibnamefont
  {{Colombi}}},\ }\href {\doibase 10.1086/174834} {\bibfield  {journal}
  {\bibinfo  {journal} {\bibapj}\ }\textbf {\bibinfo {volume} {435}},\ \bibinfo
  {pages} {536} (\bibinfo {year} {1994})},\ \Eprint
  {http://arxiv.org/abs/astro-ph/9402071} {astro-ph/9402071} \BibitemShut
  {NoStop}%
\bibitem [{\citenamefont {{Xavier}}\ \emph {et~al.}(2016)\citenamefont
  {{Xavier}}, \citenamefont {{Abdalla}},\ and\ \citenamefont
  {{Joachimi}}}]{Xavier16}%
  \BibitemOpen
  \bibfield  {author} {\bibinfo {author} {\bibfnamefont {H.~S.}\ \bibnamefont
  {{Xavier}}}, \bibinfo {author} {\bibfnamefont {F.~B.}\ \bibnamefont
  {{Abdalla}}}, \ and\ \bibinfo {author} {\bibfnamefont {B.}~\bibnamefont
  {{Joachimi}}},\ }\href@noop {} {\bibfield  {journal} {\bibinfo  {journal}
  {ArXiv e-prints}\ } (\bibinfo {year} {2016})},\ \Eprint
  {http://arxiv.org/abs/1602.08503} {arXiv:1602.08503} \BibitemShut {NoStop}%
\bibitem [{\citenamefont {{Yu}}\ \emph {et~al.}(2012)\citenamefont {{Yu}},
  \citenamefont {{Zhang}}, \citenamefont {{Lin}}, \citenamefont {{Cui}},\ and\
  \citenamefont {{Fry}}}]{yuyu12}%
  \BibitemOpen
  \bibfield  {author} {\bibinfo {author} {\bibfnamefont {Y.}~\bibnamefont
  {{Yu}}}, \bibinfo {author} {\bibfnamefont {P.}~\bibnamefont {{Zhang}}},
  \bibinfo {author} {\bibfnamefont {W.}~\bibnamefont {{Lin}}}, \bibinfo
  {author} {\bibfnamefont {W.}~\bibnamefont {{Cui}}}, \ and\ \bibinfo {author}
  {\bibfnamefont {J.~N.}\ \bibnamefont {{Fry}}},\ }\href {\doibase
  10.1103/PhysRevD.86.023515} {\bibfield  {journal} {\bibinfo  {journal}
  {\bibprd}\ }\textbf {\bibinfo {volume} {86}},\ \bibinfo {eid} {023515}
  (\bibinfo {year} {2012})},\ \Eprint {http://arxiv.org/abs/1201.4527}
  {arXiv:1201.4527 [astro-ph.CO]} \BibitemShut {NoStop}%
\bibitem [{\citenamefont {{Scherrer}}\ \emph {et~al.}(2010)\citenamefont
  {{Scherrer}}, \citenamefont {{Berlind}}, \citenamefont {{Mao}},\ and\
  \citenamefont {{McBride}}}]{Scherrer10}%
  \BibitemOpen
  \bibfield  {author} {\bibinfo {author} {\bibfnamefont {R.~J.}\ \bibnamefont
  {{Scherrer}}}, \bibinfo {author} {\bibfnamefont {A.~A.}\ \bibnamefont
  {{Berlind}}}, \bibinfo {author} {\bibfnamefont {Q.}~\bibnamefont {{Mao}}}, \
  and\ \bibinfo {author} {\bibfnamefont {C.~K.}\ \bibnamefont {{McBride}}},\
  }\href {\doibase 10.1088/2041-8205/708/1/L9} {\bibfield  {journal} {\bibinfo
  {journal} {\bibapjl}\ }\textbf {\bibinfo {volume} {708}},\ \bibinfo {pages}
  {L9} (\bibinfo {year} {2010})},\ \Eprint {http://arxiv.org/abs/0909.5187}
  {arXiv:0909.5187} \BibitemShut {NoStop}%
\bibitem [{\citenamefont {{Jing}}\ \emph {et~al.}(2007)\citenamefont {{Jing}},
  \citenamefont {{Suto}},\ and\ \citenamefont {{Mo}}}]{jingyp07}%
  \BibitemOpen
  \bibfield  {author} {\bibinfo {author} {\bibfnamefont {Y.~P.}\ \bibnamefont
  {{Jing}}}, \bibinfo {author} {\bibfnamefont {Y.}~\bibnamefont {{Suto}}}, \
  and\ \bibinfo {author} {\bibfnamefont {H.~J.}\ \bibnamefont {{Mo}}},\ }\href
  {\doibase 10.1086/511130} {\bibfield  {journal} {\bibinfo  {journal}
  {\bibapj}\ }\textbf {\bibinfo {volume} {657}},\ \bibinfo {pages} {664}
  (\bibinfo {year} {2007})},\ \Eprint {http://arxiv.org/abs/astro-ph/0610099}
  {astro-ph/0610099} \BibitemShut {NoStop}%
\bibitem [{\citenamefont {{Pan}}\ and\ \citenamefont
  {{Szapudi}}(2005)}]{panjun05}%
  \BibitemOpen
  \bibfield  {author} {\bibinfo {author} {\bibfnamefont {J.}~\bibnamefont
  {{Pan}}}\ and\ \bibinfo {author} {\bibfnamefont {I.}~\bibnamefont
  {{Szapudi}}},\ }\href {\doibase 10.1111/j.1365-2966.2005.09407.x} {\bibfield
  {journal} {\bibinfo  {journal} {\bibmnras}\ }\textbf {\bibinfo {volume}
  {362}},\ \bibinfo {pages} {1363} (\bibinfo {year} {2005})},\ \Eprint
  {http://arxiv.org/abs/astro-ph/0505422} {astro-ph/0505422} \BibitemShut
  {NoStop}%
\bibitem [{\citenamefont {{Hartlap}}\ \emph {et~al.}(2007)\citenamefont
  {{Hartlap}}, \citenamefont {{Simon}},\ and\ \citenamefont
  {{Schneider}}}]{Hartlap07}%
  \BibitemOpen
  \bibfield  {author} {\bibinfo {author} {\bibfnamefont {J.}~\bibnamefont
  {{Hartlap}}}, \bibinfo {author} {\bibfnamefont {P.}~\bibnamefont {{Simon}}},
  \ and\ \bibinfo {author} {\bibfnamefont {P.}~\bibnamefont {{Schneider}}},\
  }\href {\doibase 10.1051/0004-6361:20066170} {\bibfield  {journal} {\bibinfo
  {journal} {\bibaap}\ }\textbf {\bibinfo {volume} {464}},\ \bibinfo {pages}
  {399} (\bibinfo {year} {2007})},\ \Eprint
  {http://arxiv.org/abs/astro-ph/0608064} {astro-ph/0608064} \BibitemShut
  {NoStop}%
\bibitem [{\citenamefont {{Obreschkow}}\ \emph {et~al.}(2013)\citenamefont
  {{Obreschkow}}, \citenamefont {{Power}}, \citenamefont {{Bruderer}},\ and\
  \citenamefont {{Bonvin}}}]{Obreschkow13}%
  \BibitemOpen
  \bibfield  {author} {\bibinfo {author} {\bibfnamefont {D.}~\bibnamefont
  {{Obreschkow}}}, \bibinfo {author} {\bibfnamefont {C.}~\bibnamefont
  {{Power}}}, \bibinfo {author} {\bibfnamefont {M.}~\bibnamefont {{Bruderer}}},
  \ and\ \bibinfo {author} {\bibfnamefont {C.}~\bibnamefont {{Bonvin}}},\
  }\href {\doibase 10.1088/0004-637X/762/2/115} {\bibfield  {journal} {\bibinfo
   {journal} {\bibapj}\ }\textbf {\bibinfo {volume} {762}},\ \bibinfo {eid}
  {115} (\bibinfo {year} {2013})},\ \Eprint {http://arxiv.org/abs/1211.5213}
  {arXiv:1211.5213 [astro-ph.CO]} \BibitemShut {NoStop}%
\bibitem [{\citenamefont {{Pen}}\ \emph {et~al.}(2003)\citenamefont {{Pen}},
  \citenamefont {{Zhang}}, \citenamefont {{van Waerbeke}}, \citenamefont
  {{Mellier}}, \citenamefont {{Zhang}},\ and\ \citenamefont
  {{Dubinski}}}]{Pen03}%
  \BibitemOpen
  \bibfield  {author} {\bibinfo {author} {\bibfnamefont {U.-L.}\ \bibnamefont
  {{Pen}}}, \bibinfo {author} {\bibfnamefont {T.}~\bibnamefont {{Zhang}}},
  \bibinfo {author} {\bibfnamefont {L.}~\bibnamefont {{van Waerbeke}}},
  \bibinfo {author} {\bibfnamefont {Y.}~\bibnamefont {{Mellier}}}, \bibinfo
  {author} {\bibfnamefont {P.}~\bibnamefont {{Zhang}}}, \ and\ \bibinfo
  {author} {\bibfnamefont {J.}~\bibnamefont {{Dubinski}}},\ }\href {\doibase
  10.1086/375734} {\bibfield  {journal} {\bibinfo  {journal} {\bibapj}\
  }\textbf {\bibinfo {volume} {592}},\ \bibinfo {pages} {664} (\bibinfo {year}
  {2003})},\ \Eprint {http://arxiv.org/abs/astro-ph/0302031} {astro-ph/0302031}
  \BibitemShut {NoStop}%
\bibitem [{\citenamefont {{Jarvis}}\ \emph {et~al.}(2004)\citenamefont
  {{Jarvis}}, \citenamefont {{Bernstein}},\ and\ \citenamefont
  {{Jain}}}]{Jarvis04}%
  \BibitemOpen
  \bibfield  {author} {\bibinfo {author} {\bibfnamefont {M.}~\bibnamefont
  {{Jarvis}}}, \bibinfo {author} {\bibfnamefont {G.}~\bibnamefont
  {{Bernstein}}}, \ and\ \bibinfo {author} {\bibfnamefont {B.}~\bibnamefont
  {{Jain}}},\ }\href {\doibase 10.1111/j.1365-2966.2004.07926.x} {\bibfield
  {journal} {\bibinfo  {journal} {\bibmnras}\ }\textbf {\bibinfo {volume}
  {352}},\ \bibinfo {pages} {338} (\bibinfo {year} {2004})},\ \Eprint
  {http://arxiv.org/abs/astro-ph/0307393} {astro-ph/0307393} \BibitemShut
  {NoStop}%
\bibitem [{\citenamefont {{Fu}}\ \emph {et~al.}(2014)\citenamefont {{Fu}},
  \citenamefont {{Kilbinger}}, \citenamefont {{Erben}}, \citenamefont
  {{Heymans}}, \citenamefont {{Hildebrandt}}, \citenamefont {{Hoekstra}},
  \citenamefont {{Kitching}}, \citenamefont {{Mellier}}, \citenamefont
  {{Miller}}, \citenamefont {{Semboloni}}, \citenamefont {{Simon}},
  \citenamefont {{Van Waerbeke}}, \citenamefont {{Coupon}}, \citenamefont
  {{Harnois-D{\'e}raps}}, \citenamefont {{Hudson}}, \citenamefont {{Kuijken}},
  \citenamefont {{Rowe}}, \citenamefont {{Schrabback}}, \citenamefont
  {{Vafaei}},\ and\ \citenamefont {{Velander}}}]{fulp14}%
  \BibitemOpen
  \bibfield  {author} {\bibinfo {author} {\bibfnamefont {L.}~\bibnamefont
  {{Fu}}}, \bibinfo {author} {\bibfnamefont {M.}~\bibnamefont {{Kilbinger}}},
  \bibinfo {author} {\bibfnamefont {T.}~\bibnamefont {{Erben}}}, \bibinfo
  {author} {\bibfnamefont {C.}~\bibnamefont {{Heymans}}}, \bibinfo {author}
  {\bibfnamefont {H.}~\bibnamefont {{Hildebrandt}}}, \bibinfo {author}
  {\bibfnamefont {H.}~\bibnamefont {{Hoekstra}}}, \bibinfo {author}
  {\bibfnamefont {T.~D.}\ \bibnamefont {{Kitching}}}, \bibinfo {author}
  {\bibfnamefont {Y.}~\bibnamefont {{Mellier}}}, \bibinfo {author}
  {\bibfnamefont {L.}~\bibnamefont {{Miller}}}, \bibinfo {author}
  {\bibfnamefont {E.}~\bibnamefont {{Semboloni}}}, \bibinfo {author}
  {\bibfnamefont {P.}~\bibnamefont {{Simon}}}, \bibinfo {author} {\bibfnamefont
  {L.}~\bibnamefont {{Van Waerbeke}}}, \bibinfo {author} {\bibfnamefont
  {J.}~\bibnamefont {{Coupon}}}, \bibinfo {author} {\bibfnamefont
  {J.}~\bibnamefont {{Harnois-D{\'e}raps}}}, \bibinfo {author} {\bibfnamefont
  {M.~J.}\ \bibnamefont {{Hudson}}}, \bibinfo {author} {\bibfnamefont
  {K.}~\bibnamefont {{Kuijken}}}, \bibinfo {author} {\bibfnamefont
  {B.}~\bibnamefont {{Rowe}}}, \bibinfo {author} {\bibfnamefont
  {T.}~\bibnamefont {{Schrabback}}}, \bibinfo {author} {\bibfnamefont
  {S.}~\bibnamefont {{Vafaei}}}, \ and\ \bibinfo {author} {\bibfnamefont
  {M.}~\bibnamefont {{Velander}}},\ }\href {\doibase 10.1093/mnras/stu754}
  {\bibfield  {journal} {\bibinfo  {journal} {\bibmnras}\ }\textbf {\bibinfo
  {volume} {441}},\ \bibinfo {pages} {2725} (\bibinfo {year} {2014})},\ \Eprint
  {http://arxiv.org/abs/1404.5469} {arXiv:1404.5469} \BibitemShut {NoStop}%
\bibitem [{\citenamefont {{Guo}}\ and\ \citenamefont
  {{Jing}}(2009)}]{guohong09}%
  \BibitemOpen
  \bibfield  {author} {\bibinfo {author} {\bibfnamefont {H.}~\bibnamefont
  {{Guo}}}\ and\ \bibinfo {author} {\bibfnamefont {Y.~P.}\ \bibnamefont
  {{Jing}}},\ }\href {\doibase 10.1088/0004-637X/698/1/479} {\bibfield
  {journal} {\bibinfo  {journal} {\bibapj}\ }\textbf {\bibinfo {volume}
  {698}},\ \bibinfo {pages} {479} (\bibinfo {year} {2009})},\ \Eprint
  {http://arxiv.org/abs/0904.3200} {arXiv:0904.3200 [astro-ph.CO]} \BibitemShut
  {NoStop}%
\bibitem [{\citenamefont {{Fan}}\ \emph {et~al.}(2010)\citenamefont {{Fan}},
  \citenamefont {{Shan}},\ and\ \citenamefont {{Liu}}}]{fanzh10}%
  \BibitemOpen
  \bibfield  {author} {\bibinfo {author} {\bibfnamefont {Z.}~\bibnamefont
  {{Fan}}}, \bibinfo {author} {\bibfnamefont {H.}~\bibnamefont {{Shan}}}, \
  and\ \bibinfo {author} {\bibfnamefont {J.}~\bibnamefont {{Liu}}},\ }\href
  {\doibase 10.1088/0004-637X/719/2/1408} {\bibfield  {journal} {\bibinfo
  {journal} {\bibapj}\ }\textbf {\bibinfo {volume} {719}},\ \bibinfo {pages}
  {1408} (\bibinfo {year} {2010})},\ \Eprint {http://arxiv.org/abs/1006.5121}
  {arXiv:1006.5121} \BibitemShut {NoStop}%
\bibitem [{\citenamefont {{Liu}}\ \emph
  {et~al.}(2015{\natexlab{a}})\citenamefont {{Liu}}, \citenamefont {{Pan}},
  \citenamefont {{Li}}, \citenamefont {{Shan}}, \citenamefont {{Wang}},
  \citenamefont {{Fu}}, \citenamefont {{Fan}}, \citenamefont {{Kneib}},
  \citenamefont {{Leauthaud}}, \citenamefont {{Van Waerbeke}}, \citenamefont
  {{Makler}}, \citenamefont {{Moraes}}, \citenamefont {{Erben}},\ and\
  \citenamefont {{Charbonnier}}}]{liuxk2015}%
  \BibitemOpen
  \bibfield  {author} {\bibinfo {author} {\bibfnamefont {X.}~\bibnamefont
  {{Liu}}}, \bibinfo {author} {\bibfnamefont {C.}~\bibnamefont {{Pan}}},
  \bibinfo {author} {\bibfnamefont {R.}~\bibnamefont {{Li}}}, \bibinfo {author}
  {\bibfnamefont {H.}~\bibnamefont {{Shan}}}, \bibinfo {author} {\bibfnamefont
  {Q.}~\bibnamefont {{Wang}}}, \bibinfo {author} {\bibfnamefont
  {L.}~\bibnamefont {{Fu}}}, \bibinfo {author} {\bibfnamefont {Z.}~\bibnamefont
  {{Fan}}}, \bibinfo {author} {\bibfnamefont {J.-P.}\ \bibnamefont {{Kneib}}},
  \bibinfo {author} {\bibfnamefont {A.}~\bibnamefont {{Leauthaud}}}, \bibinfo
  {author} {\bibfnamefont {L.}~\bibnamefont {{Van Waerbeke}}}, \bibinfo
  {author} {\bibfnamefont {M.}~\bibnamefont {{Makler}}}, \bibinfo {author}
  {\bibfnamefont {B.}~\bibnamefont {{Moraes}}}, \bibinfo {author}
  {\bibfnamefont {T.}~\bibnamefont {{Erben}}}, \ and\ \bibinfo {author}
  {\bibfnamefont {A.}~\bibnamefont {{Charbonnier}}},\ }\href {\doibase
  10.1093/mnras/stv784} {\bibfield  {journal} {\bibinfo  {journal} {\bibmnras}\
  }\textbf {\bibinfo {volume} {450}},\ \bibinfo {pages} {2888} (\bibinfo {year}
  {2015}{\natexlab{a}})},\ \Eprint {http://arxiv.org/abs/1412.3683}
  {arXiv:1412.3683} \BibitemShut {NoStop}%
\bibitem [{\citenamefont {{Liu}}\ \emph
  {et~al.}(2015{\natexlab{b}})\citenamefont {{Liu}}, \citenamefont {{Petri}},
  \citenamefont {{Haiman}}, \citenamefont {{Hui}}, \citenamefont
  {{Kratochvil}},\ and\ \citenamefont {{May}}}]{liujia15}%
  \BibitemOpen
  \bibfield  {author} {\bibinfo {author} {\bibfnamefont {J.}~\bibnamefont
  {{Liu}}}, \bibinfo {author} {\bibfnamefont {A.}~\bibnamefont {{Petri}}},
  \bibinfo {author} {\bibfnamefont {Z.}~\bibnamefont {{Haiman}}}, \bibinfo
  {author} {\bibfnamefont {L.}~\bibnamefont {{Hui}}}, \bibinfo {author}
  {\bibfnamefont {J.~M.}\ \bibnamefont {{Kratochvil}}}, \ and\ \bibinfo
  {author} {\bibfnamefont {M.}~\bibnamefont {{May}}},\ }\href {\doibase
  10.1103/PhysRevD.91.063507} {\bibfield  {journal} {\bibinfo  {journal}
  {\bibprd}\ }\textbf {\bibinfo {volume} {91}},\ \bibinfo {eid} {063507}
  (\bibinfo {year} {2015}{\natexlab{b}})},\ \Eprint
  {http://arxiv.org/abs/1412.0757} {arXiv:1412.0757} \BibitemShut {NoStop}%
\bibitem [{\citenamefont {{Bond}}\ \emph {et~al.}(1998)\citenamefont {{Bond}},
  \citenamefont {{Jaffe}},\ and\ \citenamefont {{Knox}}}]{Bond98}%
  \BibitemOpen
  \bibfield  {author} {\bibinfo {author} {\bibfnamefont {J.~R.}\ \bibnamefont
  {{Bond}}}, \bibinfo {author} {\bibfnamefont {A.~H.}\ \bibnamefont {{Jaffe}}},
  \ and\ \bibinfo {author} {\bibfnamefont {L.}~\bibnamefont {{Knox}}},\ }\href
  {\doibase 10.1103/PhysRevD.57.2117} {\bibfield  {journal} {\bibinfo
  {journal} {\bibprd}\ }\textbf {\bibinfo {volume} {57}},\ \bibinfo {pages}
  {2117} (\bibinfo {year} {1998})},\ \Eprint
  {http://arxiv.org/abs/astro-ph/9708203} {astro-ph/9708203} \BibitemShut
  {NoStop}%
\bibitem [{\citenamefont {{Verde}}\ \emph {et~al.}(2003)\citenamefont
  {{Verde}}, \citenamefont {{Peiris}}, \citenamefont {{Spergel}}, \citenamefont
  {{Nolta}}, \citenamefont {{Bennett}}, \citenamefont {{Halpern}},
  \citenamefont {{Hinshaw}}, \citenamefont {{Jarosik}}, \citenamefont
  {{Kogut}}, \citenamefont {{Limon}}, \citenamefont {{Meyer}}, \citenamefont
  {{Page}}, \citenamefont {{Tucker}}, \citenamefont {{Wollack}},\ and\
  \citenamefont {{Wright}}}]{Verde03}%
  \BibitemOpen
  \bibfield  {author} {\bibinfo {author} {\bibfnamefont {L.}~\bibnamefont
  {{Verde}}}, \bibinfo {author} {\bibfnamefont {H.~V.}\ \bibnamefont
  {{Peiris}}}, \bibinfo {author} {\bibfnamefont {D.~N.}\ \bibnamefont
  {{Spergel}}}, \bibinfo {author} {\bibfnamefont {M.~R.}\ \bibnamefont
  {{Nolta}}}, \bibinfo {author} {\bibfnamefont {C.~L.}\ \bibnamefont
  {{Bennett}}}, \bibinfo {author} {\bibfnamefont {M.}~\bibnamefont
  {{Halpern}}}, \bibinfo {author} {\bibfnamefont {G.}~\bibnamefont
  {{Hinshaw}}}, \bibinfo {author} {\bibfnamefont {N.}~\bibnamefont
  {{Jarosik}}}, \bibinfo {author} {\bibfnamefont {A.}~\bibnamefont {{Kogut}}},
  \bibinfo {author} {\bibfnamefont {M.}~\bibnamefont {{Limon}}}, \bibinfo
  {author} {\bibfnamefont {S.~S.}\ \bibnamefont {{Meyer}}}, \bibinfo {author}
  {\bibfnamefont {L.}~\bibnamefont {{Page}}}, \bibinfo {author} {\bibfnamefont
  {G.~S.}\ \bibnamefont {{Tucker}}}, \bibinfo {author} {\bibfnamefont
  {E.}~\bibnamefont {{Wollack}}}, \ and\ \bibinfo {author} {\bibfnamefont
  {E.~L.}\ \bibnamefont {{Wright}}},\ }\href {\doibase 10.1086/377335}
  {\bibfield  {journal} {\bibinfo  {journal} {\bibapjs}\ }\textbf {\bibinfo
  {volume} {148}},\ \bibinfo {pages} {195} (\bibinfo {year} {2003})},\ \Eprint
  {http://arxiv.org/abs/astro-ph/0302218} {astro-ph/0302218} \BibitemShut
  {NoStop}%
\bibitem [{\citenamefont {{Takahashi}}\ \emph {et~al.}(2009)\citenamefont
  {{Takahashi}}, \citenamefont {{Yoshida}}, \citenamefont {{Takada}},
  \citenamefont {{Matsubara}}, \citenamefont {{Sugiyama}}, \citenamefont
  {{Kayo}}, \citenamefont {{Nishizawa}}, \citenamefont {{Nishimichi}},
  \citenamefont {{Saito}},\ and\ \citenamefont {{Taruya}}}]{Takahashi09}%
  \BibitemOpen
  \bibfield  {author} {\bibinfo {author} {\bibfnamefont {R.}~\bibnamefont
  {{Takahashi}}}, \bibinfo {author} {\bibfnamefont {N.}~\bibnamefont
  {{Yoshida}}}, \bibinfo {author} {\bibfnamefont {M.}~\bibnamefont {{Takada}}},
  \bibinfo {author} {\bibfnamefont {T.}~\bibnamefont {{Matsubara}}}, \bibinfo
  {author} {\bibfnamefont {N.}~\bibnamefont {{Sugiyama}}}, \bibinfo {author}
  {\bibfnamefont {I.}~\bibnamefont {{Kayo}}}, \bibinfo {author} {\bibfnamefont
  {A.~J.}\ \bibnamefont {{Nishizawa}}}, \bibinfo {author} {\bibfnamefont
  {T.}~\bibnamefont {{Nishimichi}}}, \bibinfo {author} {\bibfnamefont
  {S.}~\bibnamefont {{Saito}}}, \ and\ \bibinfo {author} {\bibfnamefont
  {A.}~\bibnamefont {{Taruya}}},\ }\href {\doibase 10.1088/0004-637X/700/1/479}
  {\bibfield  {journal} {\bibinfo  {journal} {\bibapj}\ }\textbf {\bibinfo
  {volume} {700}},\ \bibinfo {pages} {479} (\bibinfo {year} {2009})},\ \Eprint
  {http://arxiv.org/abs/0902.0371} {arXiv:0902.0371 [astro-ph.CO]} \BibitemShut
  {NoStop}%
\bibitem [{\citenamefont {{Shirasaki}}\ \emph {et~al.}(2016)\citenamefont
  {{Shirasaki}}, \citenamefont {{Takada}}, \citenamefont {{Miyatake}},
  \citenamefont {{Takahashi}}, \citenamefont {{Hamana}}, \citenamefont
  {{Nishimichi}},\ and\ \citenamefont {{Murata}}}]{Shirasaki16arxiv}%
  \BibitemOpen
  \bibfield  {author} {\bibinfo {author} {\bibfnamefont {M.}~\bibnamefont
  {{Shirasaki}}}, \bibinfo {author} {\bibfnamefont {M.}~\bibnamefont
  {{Takada}}}, \bibinfo {author} {\bibfnamefont {H.}~\bibnamefont
  {{Miyatake}}}, \bibinfo {author} {\bibfnamefont {R.}~\bibnamefont
  {{Takahashi}}}, \bibinfo {author} {\bibfnamefont {T.}~\bibnamefont
  {{Hamana}}}, \bibinfo {author} {\bibfnamefont {T.}~\bibnamefont
  {{Nishimichi}}}, \ and\ \bibinfo {author} {\bibfnamefont {R.}~\bibnamefont
  {{Murata}}},\ }\href@noop {} {\bibfield  {journal} {\bibinfo  {journal}
  {ArXiv e-prints}\ } (\bibinfo {year} {2016})},\ \Eprint
  {http://arxiv.org/abs/1607.08679} {arXiv:1607.08679} \BibitemShut {NoStop}%
\bibitem [{\citenamefont {{Takada}}\ and\ \citenamefont
  {{Hu}}(2013)}]{Takada13}%
  \BibitemOpen
  \bibfield  {author} {\bibinfo {author} {\bibfnamefont {M.}~\bibnamefont
  {{Takada}}}\ and\ \bibinfo {author} {\bibfnamefont {W.}~\bibnamefont
  {{Hu}}},\ }\href {\doibase 10.1103/PhysRevD.87.123504} {\bibfield  {journal}
  {\bibinfo  {journal} {\bibprd}\ }\textbf {\bibinfo {volume} {87}},\ \bibinfo
  {eid} {123504} (\bibinfo {year} {2013})},\ \Eprint
  {http://arxiv.org/abs/1302.6994} {arXiv:1302.6994 [astro-ph.CO]} \BibitemShut
  {NoStop}%
\bibitem [{\citenamefont {{Li}}\ \emph {et~al.}(2014)\citenamefont {{Li}},
  \citenamefont {{Hu}},\ and\ \citenamefont {{Takada}}}]{liyin14}%
  \BibitemOpen
  \bibfield  {author} {\bibinfo {author} {\bibfnamefont {Y.}~\bibnamefont
  {{Li}}}, \bibinfo {author} {\bibfnamefont {W.}~\bibnamefont {{Hu}}}, \ and\
  \bibinfo {author} {\bibfnamefont {M.}~\bibnamefont {{Takada}}},\ }\href
  {\doibase 10.1103/PhysRevD.89.083519} {\bibfield  {journal} {\bibinfo
  {journal} {\bibprd}\ }\textbf {\bibinfo {volume} {89}},\ \bibinfo {eid}
  {083519} (\bibinfo {year} {2014})},\ \Eprint {http://arxiv.org/abs/1401.0385}
  {arXiv:1401.0385} \BibitemShut {NoStop}%
\bibitem [{\citenamefont {{Mohammed}}\ and\ \citenamefont
  {{Seljak}}(2014)}]{Mohammed14}%
  \BibitemOpen
  \bibfield  {author} {\bibinfo {author} {\bibfnamefont {I.}~\bibnamefont
  {{Mohammed}}}\ and\ \bibinfo {author} {\bibfnamefont {U.}~\bibnamefont
  {{Seljak}}},\ }\href {\doibase 10.1093/mnras/stu1972} {\bibfield  {journal}
  {\bibinfo  {journal} {\bibmnras}\ }\textbf {\bibinfo {volume} {445}},\
  \bibinfo {pages} {3382} (\bibinfo {year} {2014})},\ \Eprint
  {http://arxiv.org/abs/1407.0060} {arXiv:1407.0060} \BibitemShut {NoStop}%
\bibitem [{\citenamefont {{Eifler}}\ \emph {et~al.}(2009)\citenamefont
  {{Eifler}}, \citenamefont {{Schneider}},\ and\ \citenamefont
  {{Hartlap}}}]{Eifler09}%
  \BibitemOpen
  \bibfield  {author} {\bibinfo {author} {\bibfnamefont {T.}~\bibnamefont
  {{Eifler}}}, \bibinfo {author} {\bibfnamefont {P.}~\bibnamefont
  {{Schneider}}}, \ and\ \bibinfo {author} {\bibfnamefont {J.}~\bibnamefont
  {{Hartlap}}},\ }\href {\doibase 10.1051/0004-6361/200811276} {\bibfield
  {journal} {\bibinfo  {journal} {\bibaap}\ }\textbf {\bibinfo {volume}
  {502}},\ \bibinfo {pages} {721} (\bibinfo {year} {2009})},\ \Eprint
  {http://arxiv.org/abs/0810.4254} {arXiv:0810.4254} \BibitemShut {NoStop}%
\bibitem [{\citenamefont {{Reischke}}\ \emph {et~al.}(2016)\citenamefont
  {{Reischke}}, \citenamefont {{Kiessling}},\ and\ \citenamefont
  {{Sch{\"a}fer}}}]{Reischke16}%
  \BibitemOpen
  \bibfield  {author} {\bibinfo {author} {\bibfnamefont {R.}~\bibnamefont
  {{Reischke}}}, \bibinfo {author} {\bibfnamefont {A.}~\bibnamefont
  {{Kiessling}}}, \ and\ \bibinfo {author} {\bibfnamefont {B.~M.}\ \bibnamefont
  {{Sch{\"a}fer}}},\ }\href@noop {} {\bibfield  {journal} {\bibinfo  {journal}
  {ArXiv e-prints}\ } (\bibinfo {year} {2016})},\ \Eprint
  {http://arxiv.org/abs/1607.03136} {arXiv:1607.03136} \BibitemShut {NoStop}%
\bibitem [{\citenamefont {{Taylor}}\ and\ \citenamefont
  {{Joachimi}}(2014)}]{Taylor14}%
  \BibitemOpen
  \bibfield  {author} {\bibinfo {author} {\bibfnamefont {A.}~\bibnamefont
  {{Taylor}}}\ and\ \bibinfo {author} {\bibfnamefont {B.}~\bibnamefont
  {{Joachimi}}},\ }\href {\doibase 10.1093/mnras/stu996} {\bibfield  {journal}
  {\bibinfo  {journal} {\bibmnras}\ }\textbf {\bibinfo {volume} {442}},\
  \bibinfo {pages} {2728} (\bibinfo {year} {2014})},\ \Eprint
  {http://arxiv.org/abs/1402.6983} {arXiv:1402.6983} \BibitemShut {NoStop}%
\end{thebibliography}%


\end{document}